\documentclass[10pt,a4paper,twocolumn,accepted=2021-04-12]{quantumarticle}

\pdfoutput=1

\usepackage[UKenglish]{babel}
\usepackage[UKenglish,cleanlook]{isodate}
\usepackage{amsmath}
\usepackage{amssymb}
\usepackage{amsthm}
\usepackage{IEEEtrantools}
\usepackage[bb=boondox]{mathalfa}
\usepackage{microtype}
\usepackage{alltt}
\usepackage{verbatim}
\usepackage[square,comma,numbers,sort&compress]{natbib}
\usepackage{hyperref}
\usepackage{quant_defs}

\def\bsmf{Bull.\ Soc.\ Math.\ Fr.}%
\def\hjsupp{Hadronic J.\ Suppl.}%
\def\natcomm{Nat.\ Commun.}%
\def\natphys{Nature Phys.}%
\def\njp{New J.\ Phys.}%
\def\physics{Physics}%
\def\pra{Phys.\ Rev.\ A}%
\def\prl{Phys.\ Rev.\ Lett.}%
\def\prresearch{Phys.\ Rev.\ Research}%
\def\prxquantum{PRX\ Quantum}%
\def\prsa{Proc.\ R.\ Soc.\ A}%
\def\quantic{Quantum Inf.\ Comput.}%
\def\quantum{Quantum}%
\def\qscitech{Quantum Sci.\ Technol.}%
\def\rmp{Rev.\ Mod.\ Phys.}%
\def\siamjc{SIAM J.\ Comput.}%

\newcommand{\vect}[1]{\boldsymbol{#1}}
\newcommand{\mat}[1]{\mathbf{#1}}

\newcommand{\sx}{\mathrm{X}}
\newcommand{\sy}{\mathrm{Y}}
\newcommand{\sz}{\mathrm{Z}}
\newcommand{\sv}{\vect{\sigma}}

\newcommand{\rx}{\mathrm{x}}

\newcommand{\rz}{\mathrm{z}}

\newcommand{\rxx}{\mathrm{xx}}
\newcommand{\rxz}{\mathrm{xz}}
\newcommand{\rzx}{\mathrm{zx}}
\newcommand{\rzz}{\mathrm{zz}}

\newcommand{\rA}{\mathrm{A}}
\newcommand{\rB}{\mathrm{B}}
\newcommand{\rE}{\mathrm{E}}
\newcommand{\rF}{\mathrm{F}}
\newcommand{\rAB}{\mathrm{AB}}
\newcommand{\rAE}{\mathrm{AE}}
\newcommand{\rBE}{\mathrm{BE}}
\newcommand{\rABE}{\mathrm{ABE}}

\newcommand{\guess}{\mathrm{g}}
\newcommand{\opt}{\mathrm{opt}}
\newcommand{\rmin}{\mathrm{min}}

\newcommand{\dd}{\mathrm{d}}

\newcommand{\phiA}{\varphi_{\rA}}
\newcommand{\phiB}{\varphi_{\rB}}

\newtheorem{lemma}{Lemma}

\newcommand{\liq}{Laboratoire d'Information Quantique, CP~225,
  Universit\'{e} libre de Bruxelles (ULB), \newline
  Av.\ F.~D.~Roosevelt 50, 1050 Bruxelles, Belgium}

\newcommand{\icfo}{ICFO -- Institut de Ciencies Fotoniques,
  The Barcelona Institute of Science and Technology, \newline
  08860 Castelldefels (Barcelona), Spain}

\title{Device-independent quantum key distribution \newline
  with asymmetric CHSH inequalities}

\author{Erik~Woodhead}
\affiliation{\liq}
\affiliation{\icfo}
\orcid{0000-0001-7696-4461}
\email{erik.woodhead@ulb.ac.be}

\author{Antonio~Ac\'{\i}n}
\affiliation{\icfo}
\orcid{0000-0002-1355-3435}

\author{Stefano~Pironio}
\affiliation{\liq}
\orcid{0000-0002-7196-2608}

\date{12~April~2021}

\begin{document}

\maketitle

\begin{abstract}
  The simplest device-independent quantum key distribution protocol is based
  on the Clauser-Horne-Shimony-Holt (CHSH) Bell inequality and allows two
  users, Alice and Bob, to generate a secret key if they observe sufficiently
  strong correlations. There is, however, a mismatch between the protocol, in
  which only one of Alice's measurements is used to generate the key, and the
  CHSH expression, which is symmetric with respect to Alice's two
  measurements. We therefore investigate the impact of using an extended
  family of Bell expressions where we give different weights to Alice's
  measurements. Using this family of asymmetric Bell expressions improves the
  robustness of the key distribution protocol for certain
  experimentally-relevant correlations. As an example, the tolerable error
  rate improves from $7.15\%$ to about $7.42\%$ for the depolarising
  channel. Adding random noise to Alice's key before the postprocessing
  pushes the threshold further to more than $8.34\%$. The main technical
  result of our work is a tight bound on the von Neumann entropy of one of
  Alice's measurement outcomes conditioned on a quantum eavesdropper for the
  family of asymmetric CHSH expressions we consider and allowing for an
  arbitrary amount of noise preprocessing.
\end{abstract}

\section{Introduction}

Device-independent quantum key distribution (DIQKD) allows distant parties to
create and share a cryptographic key whose security can be proved based only
on the detection of Bell-nonlocal correlations
\cite{ref:b1964,ref:bc2014,ref:ab2007}. Its signature feature is that no
assumptions are made about the quantum state and measurements performed
during the security analysis. DIQKD schemes are, correspondingly, naturally
robust against imperfections and some forms of malicious tampering with the
equipment.

The simplest protocol \cite{ref:ab2007,ref:pa2009}, inspired by a proposal by
Ekert \cite{ref:e1991}, is based around the well-known CHSH Bell inequality
\cite{ref:ch1969}. In this scheme, pairs of entangled particles are
repeatedly prepared and distributed between two parties, Alice and Bob. On a
random subset of these entangled pairs, Alice performs one out of two
$\pm 1$-valued measurements, $A_{1}$ or $A_{2}$, on the particles she
receives, and Bob similarly performs randomly one of three $\pm 1$-valued
measurements $B_{1}$, $B_{2}$, or $B_{3}$. The measurement results are used
to estimate the value of the CHSH correlator,
\begin{equation}\label{eq:chsh}
  S = \avg{A_{1} B_{1}} + \avg{A_{1} B_{2}} + \avg{A_{2} B_{1}}
  - \avg{A_{2} B_{2}} \,,
\end{equation}
as well as the value of the correlator $\avg{A_1 B_3}$, where
$\avg{A_x B_y} = P(A_{x} = B_{y}) - P(A_{x} \neq B_{y})$ and
$P(A_{x} = B_{y})$ and $P(A_{x} \neq B_{y})$ are the probability that the
outcomes of the measurements $A_{x}$ and $B_{y}$ are equal and different,
respectively. On the remaining subset of entangled particles, Alice always
performs the measurement $A_{1}$ and Bob always performs the measurement
$B_{3}$. The corresponding outcomes are then used to generate, after
classical postprocessing, a shared secret key known only to Alice and
Bob. This is possible if the estimates of the correlator $\avg{A_1 B_3}$ and
of the CHSH value are both sufficiently large. Indeed, the first condition
implies that the raw outcomes of Alice and Bob are correlated enough to be
turned into a \emph{shared} key using classical error correction. A strong
CHSH value implies, on the other hand, that their outcomes are only weakly
correlated to a potential adversary and thus that the key can be made almost
ideally \emph{secret} using privacy amplification.

This tradeoff between the CHSH expression and the adversary's knowledge,
which forms the basis of the security, can be expressed as the following
tight bound
\begin{equation}
  \label{eq:H-CHSH-bound}
  H(A_{1} | \rE)
  \geq 1 - \phi \Bigro{\sqrt{S^{2}/4 - 1}} \,,
\end{equation}
on the von Neumann entropy of Alice's outcome conditioned on an
eavesdropper's quantum side information, where
\begin{equation}
  \phi(x)
  = 1 - \tfrac{1}{2} (1 + x) \log_{2}(1 + x)
  - \tfrac{1}{2} (1 - x) \log_{2}(1 - x)
\end{equation}
is a function related to the binary entropy by
$\phi(x) = h \bigro{\tfrac{1}{2} + \tfrac{1}{2} x}$. This bound is
device-independent in that it is valid independently of the measurements
$A_{1}$, $A_{2}$, $B_{1}$, $B_{2}$ performed by Alice and Bob and the state
they share, which could be arbitrarily entangled with the adversary, under
the constraint of the expected CHSH value $S$ observed between Alice and Bob.

The bound \eqref{eq:H-CHSH-bound} is not only of fundamental interest. It has
recently been shown through the Entropy Accumulation Theorem (EAT)
\cite{ref:afd2018} (see also \cite{ref:zfk2020}) that proving unconditional
security in the finite-key regime of a DIQKD protocol consisting of $n$
measurement runs can be entirely reduced to bounding the conditional von
Neumann entropy as a function of a Bell expression, exactly as
\eqref{eq:H-CHSH-bound} does for the CHSH case.

Furthermore, a bound on the conditional von Neumann entropy directly
translates into a bound on the rate at which key bits can be generated
securely per key generation round in the asymptotic limit of many runs
$n \to \infty$. Indeed, the rates derived from the EAT approach in this
asymptotic limit (up to terms that are sublinearly decreasing in $n$) are
given by the Devetak-Winter rate \cite{ref:dw2005,ref:rgk2005}
\begin{equation}
  \label{eq:devetak-winter}
  r = H(A_{1} | \rE) - H(A_{1} | B_{3}) \,,
\end{equation}
where $H(A_{1} | B_{3})$ is the conditional Shannon entropy associated with
probabilities $P(ab|13)$ that Alice and Bob jointly obtain the outcomes $a$
and $b$ when they measure $A_{1}$ and $B_{3}$. The Devetak-Winter rate is
saturated by a class of attacks, called collective attacks, where an
eavesdropper attacks the protocol in an i.i.d.\ fashion, but where the
eavesdropper can retain quantum side information indefinitely. Inserting the
bound \eqref{eq:H-CHSH-bound} in the Devetak-Winter rate
\eqref{eq:devetak-winter} gives the tight lower bound
\begin{equation}
  \label{eq:rdw-chsh}
  r \geq 1 - \phi \Bigro{\sqrt{S^{2}/4 - 1}} - H(A_{1} | B_{3})
\end{equation}
on the asymptotic key rate for the CHSH protocol in terms of the CHSH
parameter $S$ and $H(A_{1} | B_{3})$. It is positive for sufficiently high
values of $S$ and sufficiently good correlations between the outcomes of the
measurements $A_{1}$ and $B_{3}$.

The lower bound \eqref{eq:rdw-chsh} on the Devetak-Winter rate for the
CHSH-based protocol was first presented in \cite{ref:ab2007} and derived in
detail in \cite{ref:pa2009}. The main result of \cite{ref:ab2007,ref:pa2009}
was essentially\footnote{More precisely, Ref.~\cite{ref:pa2009} derived the
  tight bound $\chi(A_{1} : \rE) \leq \phi \bigro{\sqrt{S^{2}/4 - 1}}$ on the
  Holevo quantity assuming a symmetrisation procedure is applied in the
  protocol. This was necessary in \cite{ref:pa2009} as the bound on
  $\chi(A_{1} : \rE)$ no longer generally holds if Alice's measurement
  outcomes are not equiprobable. By contrast, the analogue
  \eqref{eq:H-CHSH-bound} that we state here for the conditional von Neumann
  entropy holds generally and this will also be a feature of the more general
  bound we derive in this work.} a derivation of the bound
\eqref{eq:H-CHSH-bound} on the conditional entropy $H(A_{1} | \rE)$ through
an explicit attack saturating it (thus establishing the tightness of the
bound).

The main result presented in this paper is a tight bound on the
conditional von Neumann entropy that extends the bound
\eqref{eq:H-CHSH-bound} in two ways. First, it generalises it to the
family of CHSH-like expressions
\begin{equation}
  \label{eq:s-alpha-def}
  S_{\alpha} = \alpha \avg{A_{1} B_{1}} + \alpha \avg{A_{1} B_{2}}
  + \avg{A_{2} B_{1}} - \avg{A_{2} B_{2}} \,,
\end{equation}
where $\alpha \in \mathbb{R}$ is a parameter that can be chosen freely
($\alpha = 1$ corresponds to the regular CHSH expression). Second, it
incorporates an arbitrary level of noise preprocessing~\cite{ref:rgk2005}.

A first motivation for considering these generalisations is purely
theoretical. While we now understand how the security of a generic DIQKD
protocol can be reduced to computing bounds on the conditional von Neumann
entropy (or more precisely the derivation of what the authors of
\cite{ref:afd2018} call \emph{min-tradeoff functions}), obtaining tight or
reasonably good bounds beyond the already solved case of the CHSH expression,
the simplest Bell expression, is challenging
\cite{ref:ts2019,ref:sg2020,ref:gm2021,ref:bff2021}. Our work shows how the
von Neumann entropy can be computed for a new class of protocols and our
approach, which partly relies on reducing the problem to the well-known BB84
protocol \cite{ref:bb1984}, might inspire further, more general, results.

A second motivation is more practical. Demonstrating a working and secure
device-independent protocol remains technologically highly challenging
\cite{ref:mvd2019,ref:km2020} as it requires entangled particles to be
distributed and detected with low noise and a high detection rate over long
distances. Our results lead to two refinements to the CHSH-based protocol
that ease these demands.

The first refinement, basing the security analysis on the extended family
\eqref{eq:s-alpha-def} of Bell expressions, is motivated by the tightness of
\eqref{eq:H-CHSH-bound}. While the entropy bound \eqref{eq:H-CHSH-bound} can
be attained with equality, the eavesdropping strategy \cite{ref:ab2007} that
achieves it produces asymmetric correlations. For the optimal collective
attack, the two-body correlators in the CHSH expression are related to the
CHSH expectation value $S$ by
\begin{IEEEeqnarray}{rCl+rCl}
  \avg{A_{1} B_{1}} &=& \frac{2}{S} \,, &
  \avg{A_{1} B_{2}} &=& \frac{2}{S} \,, \\
  \avg{A_{2} B_{1}} &=& \frac{S^{2}/2 - 2}{S} \,, &
  \avg{A_{2} B_{2}} &=& -\frac{S^{2}/2 - 2}{S} \,. \IEEEeqnarraynumspace
\end{IEEEeqnarray}
This reflects an asymmetry in the protocol: Alice uses the $A_{1}$
measurement to generate the key while $A_{2}$ is only used for parameter
estimation. To mitigate this, instead of using only CHSH we will consider the
extended family of Bell expressions \eqref{eq:s-alpha-def} where a different
weight $\alpha \in \mathbb{R}$ is given to the correlation terms involving
$A_{1}$.

Bounding the conditional entropy for the family \eqref{eq:s-alpha-def} and
then choosing whichever value of $\alpha$ gives the highest result amounts to
the same as bounding the conditional entropy in terms of the combinations
$\avg{A_{1} B_{1}} + \avg{A_{1} B_{2}}$ and
$\avg{A_{2} B_{1}} - \avg{A_{2} B_{2}}$ viewed as independent parameters. In
general, it has been observed that using more information about the
statistics can improve the performance of a device-independent cryptography
protocol \cite{ref:nsps2014,ref:bss2014}.

The second refinement, noise preprocessing, consists of a classical change to
the protocol in which Alice randomly flips each of her key bits intended for
key generation with some probability $q$, known publicly, before the
classical postprocessing to distil the secret key is applied. Noise
preprocessing is known to improve the robustness of QKD protocols
\cite{ref:rgk2005}. Intuitively, adding random noise to Alice's outcomes
makes things worse (increases $H(A_{1} | B_{3})$) for Alice and Bob, but it
also makes things worse (increases $H(A_{1} | \rE)$) for the eavesdropper and
it turns out the result can be a net increase to the key rate.

Both refinements are simply incorporated to the standard DIQKD protocol of
\cite{ref:ab2007} given our generalisation of the conditional entropy bound
\eqref{eq:H-CHSH-bound} for the family $S_{\alpha}$ of Bell expressions with
noise preprocessing. As we will see in our case, deriving the entropy bound
essentially reduces to deriving the conditional entropy bound for the
well-known BB84 \cite{ref:bb1984} QKD protocol. We give a short outline of
how this works for the entropy bound \eqref{eq:H-CHSH-bound} for CHSH in
section~\ref{sec:simple-chsh} before giving the full derivation of our main
result in section~\ref{sec:derivation}. We then derive some examples of its
effect on the robustness of the DIQKD protocol in
section~\ref{sec:keyrate-examples}.

\section{The entropy bound}
\label{sec:main-result}

Let Alice, Bob, and an adversary, Eve, share some arbitrary tripartite state
$\rho_{\rABE}$, and let $A_{1}$ and $A_{2}$ be two arbitrary
binary-valued\footnote{In the following, we freely switch back and forth from
  a description where Alice's and Bob's measurement results take the values
  $\{0,1\}$ or the values $\{+1,-1\}$. This is just a convention and the
  choice depends on what is more convenient in terms of notation.}
measurements on Alice's system and $B_{1}$ and $B_2$ two arbitrary
binary-valued measurements on Bob's system. We can think of the state and
measurements as chosen by Eve.  Without loss of generality we may assume the
measurements to be projective (if necessary by increasing the Hilbert space
dimensions).

If Alice measures $A_{1}$ and flips her outcome with probability
$q \in [0,1]$, the correlations between Alice and Eve are described by the
classical-quantum state
\begin{equation}
  \label{eq:tau_ZE_1}
  \tau_{\rAE}
  = [0]_{\rA} \otimes (\bar{q} \rho_{\rE}^{0} + q \rho_{\rE}^{1})
  + [1]_{\rA} \otimes (q \rho_{\rE}^{0} + \bar{q} \rho_{\rE}^{1}) \,,
\end{equation}
where $\bar{q} = 1 - q$, $[0]$ and $[1]$ are shorthand for classical register
states $\proj{0}$ and $\proj{1}$, and
\begin{equation}
\rho_{\rE}^{a} = \Tr_{\rAB}[\Pi_{a} \rho_{\rABE}] \,,
\end{equation}
where $\Pi_{0,1} = (\id \pm A_{1}) / 2$ are the projectors associated with
Alice's $A_{1}$ measurement. The conditional entropy of Alice's final outcome
conditioned on Eve's knowledge is then defined as
\begin{equation}\label{eq:cond-entr}
  H(A_{1} | \rE) = S(\tau_{\rAE}) - S(\tau_{\rE})
\end{equation}
where
$\tau_{\rE} = \Tr_{\rA}[\tau_{\rAE}] = \sum_{a} \rho_{\rE}^{a} = \rho_{\rE}$
is Eve's average reduced state, $S(\rho) = -\Tr[\rho \log_{2}(\rho)]$ is the
von Neumann entropy, and $\log_{2}$ is the logarithm function in base 2.

The main result that we derive is a family of lower bounds
\begin{equation}
  H(A_{1} | \rE) \geq \bar{g}_{q,\alpha}(S_{\alpha})
\end{equation}
on the conditional von Neumann entropy in terms of the expectation value
\eqref{eq:s-alpha-def} of the Bell expression $S_{\alpha}$ computed on the
reduced state $\rho_{\rAB}=\Tr_{\rE}[\rho_{\rABE}]$, valid for any values of
the parameters $\alpha \in \mathbb{R}$ and $ q \in [0,1]$. These bounds hold
for any state $\rho_{\rABE}$ and measurements $A_{1}$, $A_{2}$, $B_{1}$,
$B_{2}$ and are hence device-independent.

\begin{figure}[tbp]
  \centering
  \includegraphics{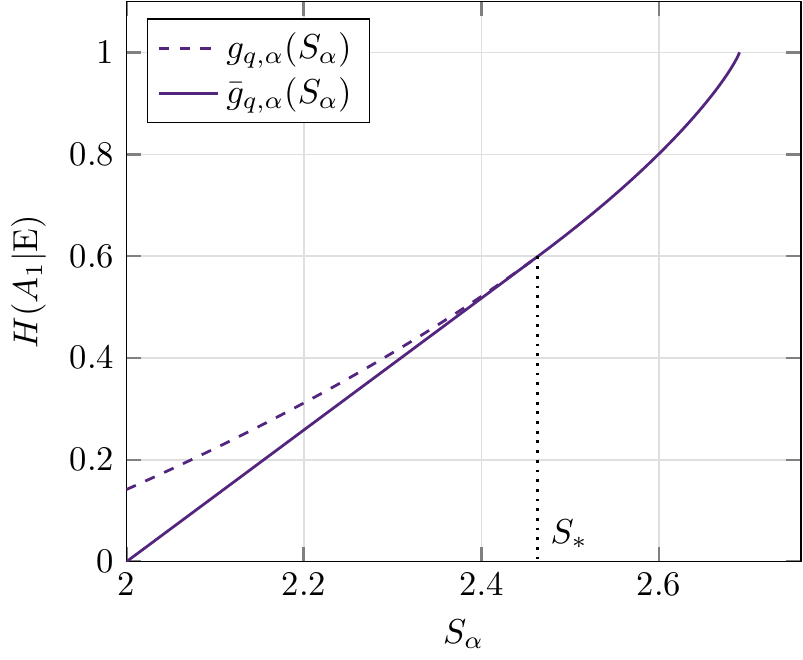}
  \caption{Conditional von Neumann entropy $H(A_{1} | \rE)$ as a function on
    the observed value of $S_{\alpha}$ given by our explicit attack,
    illustrated here for $q = 0$ and $\alpha = 0.9$, which is representative
    for values $\abs{\alpha} < 1$. The dashed line is a plot of
    \eqref{eq:g-qalpha}. It is visibly too high to be the optimal
    device-independent strategy for all $S_{\alpha}$ given that the real
    curve must be convex and attain $h(q) = 0$ at the classical bound
    $S_{\alpha} = 2$. To get the correct relation, we use the tangent of
    $g_{q,\alpha}$ for values of $S_{\alpha}$ less than the point $S_{*}$
    where the tangent intersects the point
    $\bigro{H(A_{1} | \rE),\, S_{\alpha}} = \bigro{h(q), 2}$. For $q = 0$ and
    $\alpha = 0.9$ this happens at $S_{*} \approx 2.4634$.}
  \label{fig:HA_E-Salpha}
\end{figure}

The function $\bar{g}_{q,\alpha}$ is piecewise defined and its construction
is described below and illustrated for $q = 0$ and $\alpha = 0.9$ in
figure~\ref{fig:HA_E-Salpha}. As a way of explaining its form, we introduce
it via a strategy that we considered as a candidate for the optimal
collective attack.

The strategy is a minor modification of the optimal attack
\citep{ref:ab2007,ref:pa2009} saturating the CHSH bound
\eqref{eq:H-CHSH-bound}. Eve prepares a pure tripartite state
$\rho_{\rABE}=\proj{\Psi_\rABE}$ of the form
\begin{equation}
  \label{eq:PsiABE-optattack}
  \ket{\Psi}_{\rABE} = \frac{1}{\sqrt{2}}
  \Bigro{\ket{00}_{\rAB} \ket{\psi_{0}}_{\rE}
    + \ket{11}_{\rAB} \ket{\psi_{1}}_{\rE}}\,,
\end{equation}
where the strength of the
attack is determined by the overlap
\begin{equation}
  \braket{\psi_{0}}{\psi_{1}} = F \in [0, 1]\,.
\end{equation}
Alice and Bob then measure
\begin{IEEEeqnarray}{rCl+rCl}
  A_{1} &=& \sz \,, & A_{2} &=& \sx
\end{IEEEeqnarray}
and
\begin{IEEEeqnarray}{rCl}
  B_{1} &=& \cos \bigro{\tfrac{\phiB}{2}} \sz
  + \sin \bigro{\tfrac{\phiB}{2}} \sx \,, \\
  B_{2} &=& \cos \bigro{\tfrac{\phiB}{2}} \sz
  - \sin \bigro{\tfrac{\phiB}{2}} \sx \,,
\end{IEEEeqnarray}
where $\sz$ and $\sx$ are the eponymous Pauli operators and $\phiB$ is an
angle that we will optimise momentarily. The classical-quantum state after
Alice measures $A_{1}$ and flips her outcome with probability $q$ is thus
given by \eqref{eq:tau_ZE_1} with $\rho_{\rE}^{a} = \psi_{a} / 2$, where
$\psi_{a}$ is a shorthand for $\proj{\psi_{a}}$. The conditional entropy
\eqref{eq:cond-entr} can then directly be computed in terms of the overlap
$F$ to be
\begin{equation}
  \label{eq:HAE-F-optattack}
  H(A_{1} | \rE)
  = 1 + \phi \bigro{\sqrt{(\bar{q} - q)^{2} + 4 q \bar{q} F^{2}}}
  - \phi(F) \,.
\end{equation}
On the other hand, the marginal state of Alice and Bob is
\begin{equation}
  \rho_{\rAB} = \frac{1}{4} \Bigsq{
    \id \otimes \id  + \sz \otimes \sz
    + F \bigro{\sx \otimes \sx - \sy \otimes \sy}} \,.
\end{equation}
For the above measurements and choosing an optimal angle $\phiB$ that
maximises the expectation value of $S_{\alpha}$, we find
\begin{IEEEeqnarray}{rCl}
  S_{\alpha} &=& 2 \alpha \cos \bigro{\tfrac{\phiB}{2}}
  + 2 F \sin \bigro{\tfrac{\phiB}{2}} \nonumber \\
  &=& 2 \sqrt{\alpha^{2} + F^{2}} \,,
\end{IEEEeqnarray}
which rearranges for $F$ to
\begin{equation}
  \label{eq:F-Salpha-optattack}
  F = \sqrt{S\du{\alpha}{2}/4 - \alpha^{2}}\,.
\end{equation}
Substituting \eqref{eq:F-Salpha-optattack} into
\eqref{eq:HAE-F-optattack}, we find that the conditional entropy is related
to $S_{\alpha}$ for the particular strategy we have described by
\begin{equation}
  H(A_{1} | \rE) = g_{q,\alpha}(S_{\alpha}) \,,
\end{equation}
where
\begin{IEEEeqnarray}{rCl}
  \label{eq:g-qalpha}
  g_{q,\alpha}(s) &=& 1 + \phi \Bigro{\sqrt{(1 - 2 q)^{2}
      + 4 q (1 - q) (s^{2} / 4 - \alpha^{2})}} \nonumber \\
  &&-\> \phi \Bigro{\sqrt{s^{2} / 4 - \alpha^{2}}} \,.
\end{IEEEeqnarray}

A little consideration shows that the above strategy cannot be the optimal
one minimising the entropy in all cases. The Bell expression $S_{\alpha}$ has
the classical and quantum bounds \cite{ref:llp2010,ref:amp2012}
\begin{IEEEeqnarray}{c+t+c}
  C_{\alpha} = \left\{
  \begin{IEEEeqnarraybox}[\IEEEeqnarraystrutmode
        \IEEEeqnarraystrutsizeadd{2pt}{2pt}][c]{ll}
    2 \abs{\alpha} &\text{ if } \abs{\alpha} \geq 1 \\
    2 &\text{ if } \abs{\alpha} \leq 1
  \end{IEEEeqnarraybox} \right.
  &and&
  Q_{\alpha} = 2 \sqrt{1 + \alpha^{2}} \,. \IEEEeqnarraynumspace
\end{IEEEeqnarray}
At the quantum maximum $S_{\alpha} = Q_{\alpha}$ we find
$g_{q,\alpha}(Q_\alpha) = 1$, i.e., the eavesdropper has no knowledge
whatsoever about Alice's outcome as we would naturally expect for any
conceivable strategy.

At the classical boundary $S_{\alpha} = C_{\alpha}$, we would expect an
optimal attack to yield $H(A_{1} | \rE) = h(q)$ since Alice and Bob's
correlations can be attained with a deterministic strategy and the only
randomness in Alice's outcome then comes from the noise preprocessing.  The
function \eqref{eq:g-qalpha} attains
\begin{equation}
  g_{q,\alpha}(S_{\alpha}) = h(q)
\end{equation}
at $S_{\alpha} = 2 \abs{\alpha}$. If $\abs{\alpha} \geq 1$, this is the same
as the classical bound and there is no problem. However, if
$\abs{\alpha} < 1$ then the classical bound is $C_{\alpha} = 2$ and the value
of $g_{q,\alpha}(S_{\alpha})$ at $S_{\alpha} = 2$ is too high to describe the
optimal strategy. However, we can improve it by taking probabilistic mixtures
of the above strategy with the classical one achieving
$H(A_{1} | \rE) = h(q)$ at $S_{\alpha} = 2$. Geometrically we are
considering, in the plane $\bigro{S_{\alpha}, H(A_{1} | \rE)}$, the convex
hull of the points $\bigro{S_{\alpha}, g_{q,\alpha}(S_{\alpha})}$ and
$\bigro{2, h(q)}$. As illustrated in figure~\ref{fig:HA_E-Salpha}, this
amounts to extending the curve $g_{q,\alpha}(s)$ linearly from the point
where its tangent intersects $H(A_{1} | \rE) = h(q)$ at $S_{\alpha} = 2$.

Our main result, which we prove in section~\ref{sec:derivation}, is that the
explicit attack that we just described is optimal. That is, the construction
shown in figure~\ref{fig:HA_E-Salpha} gives the device-independent lower
bound on the conditional entropy for all $\abs{\alpha} < 1$ while the bound
is simply given by $g_{q,\alpha}(S_{\alpha})$ for $\abs{\alpha} \geq 1$.

\paragraph{Main result.} To summarise in more mathematical terms, our main
result is that the conditional von Neumann entropy, computed on the
post-measurement classical-quantum state \eqref{eq:tau_ZE_1} following an
amount $q$ of noise preprocessing, is bounded in terms of $S_{\alpha}$ by
\begin{equation}
  \label{eq:HAE-bound}
  H(A_{1} | \rE) \geq \bar{g}_{q,\alpha}(S_{\alpha}) \,,
\end{equation}
where $\bar{g} \equiv \bar{g}_{q,\alpha}$ is defined in terms of
\begin{IEEEeqnarray}{rCl}
  g(s) &=& 1 + \phi \Bigro{\sqrt{(1 - 2 q)^{2}
      + 4 q (1 - q) (s^{2} / 4 - \alpha^{2})}} \nonumber \\
  &&-\> \phi \Bigro{\sqrt{s^{2} / 4 - \alpha^{2}}}
\end{IEEEeqnarray}
as
\begin{equation}
  \bar{g}(s) = \left\{
    \begin{IEEEeqnarraybox}[\IEEEeqnarraystrutmode
        \IEEEeqnarraystrutsizeadd{2pt}{2pt}][c]{ll}
      g(s)
      & \text{ if } \abs{\alpha} \geq 1
      \text{ or } s \geq s_{*} \\
      h(q) + g'(s_{*}) (\abs{s} - 2)
      & \text{ if } \abs{\alpha} < 1 \text{ and } s < s_{*}
  \end{IEEEeqnarraybox} \right. \,,
\end{equation}
where in turn $g' \equiv g'_{q,\alpha}$ is the first derivative of
$g \equiv g_{q,\alpha}$ and, for $\abs{\alpha} < 1$,
$s_{*} \equiv s_{*}(q, \alpha)$ is the unique point where the tangent of
$g(s)$ crosses $h(q)$ at $s = 2$, i.e., such that
\begin{equation}
  \label{eq:s*-root-problem}
  h(q) + g'(s_{*}) (s_{*} - 2) = g(s_{*}) \,.
\end{equation}

We note that it is sufficient to consider $s_{*}$ in the range
\begin{equation}
  \label{eq:s*-root-range}
  2 \sqrt{1 + \alpha^{2} - \alpha^{4}}
  \leq s_{*}
  \leq 2 \sqrt{1 + \alpha^{2}}\,.
\end{equation}
The upper bound corresponds to the quantum maximal value; the origin of the
lower bound will be explained at the end of section~\ref{sec:derivation}.
The attack strategy we started with shows that the entropy bound
\eqref{eq:HAE-bound} is tight and can be attained for any values of the
parameters $q$ and $\alpha$.

\begin{figure}[htbp]
  \centering
  \includegraphics{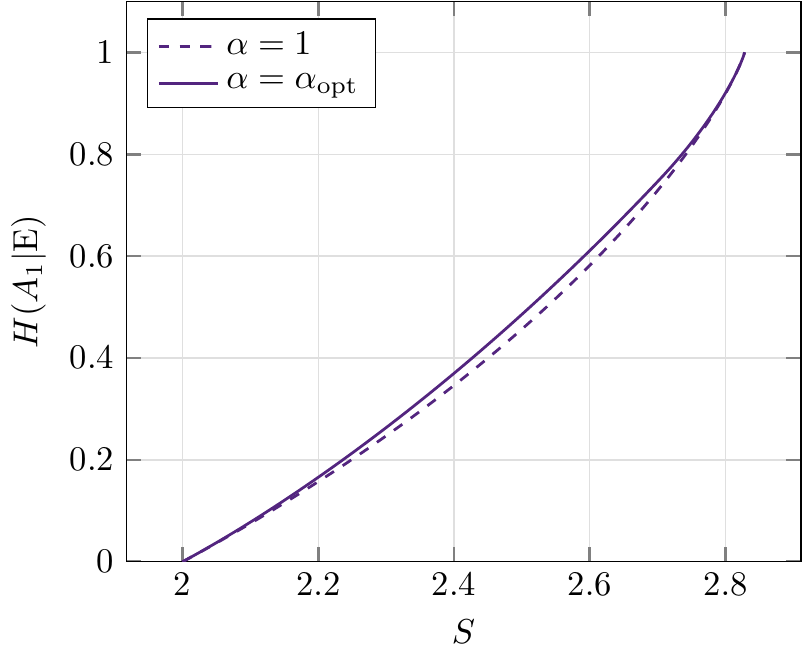}
  \caption{Lower bound on the conditional von Neumann entropy in terms of the
    CHSH expectation value using the CHSH entropy bound
    \eqref{eq:H-CHSH-bound} (dashed line) and the bound \eqref{eq:HAE-bound}
    for the $S_{\alpha}$ family for $q = 0$ and the optimal value of $\alpha$
    (solid line) for correlations satisfying
    $\avg{A_{1} B_{1}} + \avg{A_{1} B_{2}} = \avg{A_{2} B_{1}} - \avg{A_{2}
      B_{2}} = S/2$. The optimal value of $\alpha$ decreases from $1$ to
    about $0.84$ as $S$ goes from $2$ to about $2.7$ and then increases back
    to $1$ again as $S$ approaches $2 \sqrt{2}$.}
  \label{fig:HA_E-comparison}
\end{figure}

For given correlations, $\bar{g}_{q,\alpha}(S_{\alpha})$ can be maximised
over $\alpha$ to obtain the best bound on the conditional entropy in terms of
$\avg{A_{1} B_{1}} + \avg{A_{1} B_{2}}$ and
$\avg{A_{2} B_{1}} - \avg{A_{2} B_{2}}$ seen as separate parameters. The
result for $q = 0$ and correlations satisfying
\begin{equation}\label{eq:chsh-corr}
  \avg{A_{1} B_{1}} + \avg{A_{1} B_{2}}
  = \avg{A_{2} B_{1}} - \avg{A_{2} B_{2}}
  = S/2
\end{equation}
is shown and compared with the CHSH entropy bound \eqref{eq:H-CHSH-bound} in
figure~\ref{fig:HA_E-comparison}.

\section{Short derivation for CHSH}
\label{sec:simple-chsh}

In the special case of the CHSH expression ($\alpha = 1$) and that no noise
preprocessing is applied ($q = 0$), the von Neumann entropy bound
\eqref{eq:HAE-bound} and main result of this paper simplifies to
\begin{equation}
  \label{eq:HAE-CHSH-bound}
  H(A_{1} | \rE)
  \geq 1 - \phi \Bigro{\sqrt{S^{2}/4 - 1}} \,.
\end{equation}
Before proving the main result \eqref{eq:HAE-bound} we give a short
derivation here for the special case \eqref{eq:HAE-CHSH-bound}. We do this
partly just to show that there is a much simpler way to derive
\eqref{eq:HAE-CHSH-bound} than the approach originally followed in
\cite{ref:pa2009}; it also can serve as an outline for the full derivation of
\eqref{eq:HAE-bound} that we undertake in section~\ref{sec:derivation}. The
derivation is a simplified version\footnote{In terms of the notation and
  basis choices we use in this section, \cite{ref:wp2015} essentially did the
  prepare-and-measure analogue of deriving
  $\abs{\avg{\sy \otimes \sy}} \geq \sqrt{S^{2}/4 - 1}$ and combining this
  with the BB84 bound
  $H_{\rmin}(\sz | \rE) \geq 1 - \log_{2} \bigro{1 + \sqrt{1 - \avg{\sy
        \otimes \sy}^{2}}}$ for the min-entropy. Ref.~\cite{ref:wp2015}
  concentrated on bounding the min-entropy due to a complication called
  ``basis dependence'' specific to the prepare-and-measure setting that makes
  it more difficult to tightly bound the conditional von Neumann entropy in
  that case. Some results for the conditional von Neumann entropy under
  different assumptions are nevertheless presented for this setting as
  Eqs.~(4.33) and (4.36) in section~4.4 of \cite{ref:w2014b}. They can be
  generalised to incorporate noise preprocessing by using Eq.~(3.22) in place
  of Eq.~(3.14) there to obtain the final entropy bound.} of one done in
\cite{ref:w2014b,ref:wp2015} for a prepare-and-measure version of the
CHSH-based protocol.

The main idea is that we can reduce deriving \eqref{eq:HAE-CHSH-bound} to
bounding the conditional entropy for the well-known BB84 protocol
\cite{ref:bb1984}. To do this, we exploit two facts that are by now well
established for this problem: first, we can assume without
loss of generality that Alice's and Bob's measurements are projective and,
second, since both parties perform only two dichotomic measurements to
estimate CHSH, we can use the Jordan lemma to reduce the analysis to qubit
systems.

Concentrating on qubit systems, then, we know from security analyses of the
BB84 protocol (see e.g.\ \cite{ref:bc2010} or \cite{ref:w2016,ref:w2013})
that the conditional entropy of the outcome of a Pauli $\sz$ measurement by
Alice is lower bounded by
\begin{equation}
  \label{eq:HZE-bb84}
  H(\sz | \rE) \geq 1 - \phi \bigro{\abs{\avg{\sx \otimes \sx}}}
\end{equation}
in terms of the correlation $\avg{\sx \otimes \sx}$ between the outcomes of
Pauli $\sx$ measurements performed by Alice and Bob on the same initial
state. To apply \eqref{eq:HZE-bb84} to the device-independent protocol we
need to identify Alice's measurement $A_{1}$ with $\sz$. Since we assume the
measurements are projective this is straightforward to justify: the CHSH
inequality cannot be violated if any of the measurements are degenerate
(i.e., $\pm \id$) and thus must all be linear combinations of the three Pauli
operators. The only basis-independent properties characterising the
measurements then are the angles between them on the Bloch sphere. We can
therefore choose the local bases in such a way that
\begin{IEEEeqnarray}{rCl}
  A_{1} &=& \sz \,, \\
  A_{2} &=& \cos(\phiA) \sz + \sin(\phiA) \sx
\end{IEEEeqnarray}
and
\begin{IEEEeqnarray}{rCl}
  B_{1} + B_{2} &=& 2 \cos\bigro{\tfrac{\phiB}{2}} \sz \,, \\
  B_{1} - B_{2} &=& 2 \sin\bigro{\tfrac{\phiB}{2}} \sx \,,
\end{IEEEeqnarray}
where $\phiA$ and $\phiB$ are unknown angles. With this choice of bases, the
CHSH expectation value can be expressed as and then bounded by
\begin{IEEEeqnarray}{rCl}
  \label{eq:CHSH-XX-bound}
  S &=& \avg{A_{1} (B_{1} + B_{2})} + \avg{A_{2} (B_{1} - B_{2})}
  \nonumber \\
  &=& 2 \cos \bigro{\tfrac{\phiB}{2}} \avg{\sz \otimes \sz}
  + 2 \cos(\phiA)
  \sin \bigro{\tfrac{\phiB}{2}} \avg{\sz \otimes \sx}
  \nonumber \\
  &&+\> 2 \sin(\phiA)
  \sin \bigro{\tfrac{\phiB}{2}} \avg{\sx \otimes \sx}
  \nonumber \\
  &\leq& 2 \sqrt{\avg{\sz \otimes \sz}^{2}
    + \avg{\sz \otimes \sx}^{2} + \avg{\sx \otimes \sx}^{2}} \nonumber \\
  &\leq& 2 \sqrt{1 + \avg{\sx \otimes \sx}^{2}} \,,
\end{IEEEeqnarray}
where we used the Cauchy-Schwarz inequality and that
\begin{IEEEeqnarray}{rCl}
  \cos\bigro{\tfrac{\phiB}{2}}^{2}
  + \bigsq{\cos(\phiA) \sin \bigro{\tfrac{\phiB}{2}}}^{2} && \nonumber \\
  +\> \bigsq{\sin(\phiA) \sin \bigro{\tfrac{\phiB}{2}}}^{2}
  &=& 1
\end{IEEEeqnarray}
to get to the third line and a constraint
\begin{IEEEeqnarray}{rCl}
  \label{eq:ZZ-ZX-bound}
  \avg{\sz \otimes \sz}^{2} + \avg{\sz \otimes \sx}^{2} \leq 1
\end{IEEEeqnarray}
respected by correlations between Pauli operators to get to the fourth. The
inequality \eqref{eq:CHSH-XX-bound} rearranges to a lower bound
\begin{equation}
  \label{eq:XX-CHSH-bound}
  \abs{\avg{\sx \otimes \sx}} \geq
  \sqrt{S^{2} / 4 - 1}
\end{equation}
for the absolute value $\abs{\avg{\sx \otimes \sx}}$ of the correlator
appearing in the BB84 entropy bound \eqref{eq:HZE-bb84}. Since we chose the
bases in such a way as to identify $A_{1}$ with $\sz$, we simply substitute
\eqref{eq:XX-CHSH-bound} into \eqref{eq:HZE-bb84} to obtain
\eqref{eq:HAE-CHSH-bound}. The convexity of the result in $S$ then allows the
qubit bound to be extended to arbitrary dimension through Jordan's lemma.

\section{Derivation of main result}
\label{sec:derivation}

The short derivation for CHSH above illustrates the general approach and
kinds of technical ingredients we will work with to obtain a proof of the
main result \eqref{eq:HAE-bound}. A summary of the key steps is:
\begin{itemize}
  \item We reduce the problem to one where Alice's and Bob's subsystems are
  qubits.
  \item We need a generalisation of the BB84 entropy bound
  \eqref{eq:HZE-bb84} allowing for noise preprocessing ($q \neq 0$).
  \item We derive constraints on correlations between Pauli
  operators that we can work with, such as \eqref{eq:ZZ-ZX-bound}, in order
  to transform the $S_{\alpha}$ family of Bell expressions into a bound for a
  correlator $\abs{\avg{\sx \otimes B}}$ that we can use in the BB84
  entropy bound.
  \item Finally, we should determine whether the resulting qubit bound is
  convex and, if it is not, take its convex hull to obtain the fully
  device-independent bound.
\end{itemize}
While we focus here on the family of $S_{\alpha}$ expressions, for which we
are able to bound the conditional entropy analytically, we remark that our
approach of reducing the problem to qubits and then using the entropy bound
for the BB84 protocol applies generically to the two-input/two-output
device-independent setting.

\subsection{Reduction to qubits}

Following the approach used in many studies of the CHSH Bell scenario, we
start by reducing the problem to one where Alice and Bob perform qubit
measurements. We recapitulate how this works here.

The reduction is based on the Jordan lemma~\cite{ref:j1875}, which tells us
that any pair $A_{1}$, $A_{2}$ of observable operators whose eigenvalues are
all $\pm 1$ admit a common block diagonalisation in blocks of dimension no
larger than two. That is, there is a choice of bases in which the observables
appearing in the $S_{\alpha}$ Bell expression can be expressed as
\begin{IEEEeqnarray}{rCl+rCl}
  A_{x} &=& \sum_{j} A_{x|j} \otimes [j]_{\rA'} \,, & x &=& 1,2 \,, \\
  B_{y} &=& \sum_{k} B_{y|k} \otimes [k]_{\rB'} \,, & y &=& 1,2
\end{IEEEeqnarray}
for qubit operators $A_{x|j}$ and $B_{y|k}$\footnote{For simplicity we ignore
  possible $1 \times 1$ Jordan blocks; any such blocks can be grouped
  together into larger $2 \times 2$ blocks. Since the analysis we intend to
  perform is also device-independent we can also assume Alice's and Bob's
  Hilbert spaces are of even dimension without loss of generality, extending
  them if necessary.}. Proofs of this result can be found in
\cite{ref:t1993,ref:m2006,ref:pa2009}.

After Alice measures $A_{1}$ and flips her outcome with probability $q$, we
remind that the correlation between Alice and Eve is described by the
classical-quantum state
\begin{equation}
  \label{eq:tau_ZE_2}
  \tau_{\rAE}
  = [0]_{\rA} \otimes (\bar{q} \rho_{\rE}^{0} + q \rho_{\rE}^{1})
  + [1]_{\rA} \otimes (q \rho_{\rE}^{0} + \bar{q} \rho_{\rE}^{1}) \,,
\end{equation}
where
\begin{equation}
  \rho_{\rE}^{a} = \Tr_{\rAB}[\Pi_{a} \rho_{\rABE}]
\end{equation}
and $\Pi_{0,1} = (\id \pm A_{1}) / 2$ are the projectors associated with
Alice's $A_{1}$ measurement. Introducing the block diagonalisation, we can
reexpress $\rho_{\rE}^{a}$ as
\begin{equation}
  \rho_{\rE}^{a} = \sum_{jk} p_{jk} \rho^{a}_{jk}
\end{equation}
where\footnote{Here $\rA'$ and $\rB'$ are subsystems, respectively, of $\rA$
  and $\rB$. When taking the product of operators such as
  $[jk]_{\rA'\rB'} \, \rho_{\rABE}$, we omit in the notation identity
  operators on unspecified subsystems. That is,
  $[jk]_{\rA'\rB'} \, \rho_{\rABE} = \bigro{\id_{\bar\rA}\otimes [j]_{\rA'}
    \otimes \id_{\bar{\rB}} \otimes [j]_{\rB'} \otimes \id_{\rE}} \,
  \rho_{\rABE}$, where $\bar{\rA} = \rA \setminus \rA'$ and
  $\bar{\rB} = \rB \setminus \rB'$.}
\begin{IEEEeqnarray}{rCl}
  p_{jk} &=& \Tr \bigsq{[jk]_{\rA'\rB'} \, \rho_{\rABE}} \,, \\
  p_{jk} \, \rho_{jk}^{a}
  &=& \Tr_{\rAB} \bigsq{(\Pi_{a} \otimes [jk]_{\rA'\rB'}) \rho_{\rABE}} \,.
\end{IEEEeqnarray}
This allows us to reexpress $\tau_{\rAE}$ as
\begin{equation}
  \tau_{\rAE} = \sum_{jk} p_{jk} \, \tau_{jk} \,,
\end{equation}
where
\begin{equation}
  \label{eq:tau_ZE_jk}
  \tau_{jk}
  = [0]_{\rA} \otimes (\bar{q} \rho_{jk}^{0} + q \rho_{jk}^{1})
  + [1]_{\rA} \otimes (q \rho_{jk}^{0} + \bar{q} \rho_{jk}^{1}) \,.
\end{equation}

The expectation value of $S_{\alpha}$ similarly decomposes according to
\begin{equation}
  S_{\alpha} = \sum_{jk} p_{jk} S_{\alpha | jk}
\end{equation}
where $S_{\alpha | jk}$ is the contribution to $S_{\alpha}$ from the pair
$(j, k)$ of Jordan blocks. Importantly, the expectation value
$S_{\alpha | jk}$ and classical-quantum state $\tau_{jk}$ conditioned on the
Jordan blocks are both determined by the same conditional state
\begin{equation}
  p_{jk} \, \rho_{\rABE | jk}
  = \Tr_{\rA'\rB'} \bigsq{[jk]_{\rA'\rB'} \, \rho_{\rABE}}
\end{equation}
where Alice's and Bob's subsystems are qubits. This allows us to reduce the
entire problem to qubit systems. More precisely, suppose we have derived a
lower bound
\begin{equation}
  H(A_{1} | \rE) \geq \bar{g}(S_{\alpha})
\end{equation}
for the conditional entropy for qubit systems that is convex\footnote{If we
  have a bound that is not convex we take its convex hull.}. Then, concavity
of the conditional von Neumann entropy and the convexity of $\bar{g}$ imply
in arbitrary dimension
\begin{IEEEeqnarray}{rCl}
\label{eq:convex_ext}
  H(A_{1} | \rE)_{\tau}
  &\geq& \sum_{jk} p_{jk} \, H(A_{1} | \rE)_{\tau_{jk}} \nonumber \\
  &\geq& \sum_{jk} p_{jk} \, \bar{g}(S_{\alpha | jk}) \nonumber \\
  &\geq& \bar{g} \Bigro{\sum_{jk} p_{jk} \, S_{\alpha | jk}} \nonumber \\
  &=& \bar{g}(S_{\alpha})\,.
\end{IEEEeqnarray}

\subsection{BB84 entropy bound}
\label{sec:bb84-noise-bound}

We now derive the required BB84 entropy bound including noise
preprocessing. The result we derive here is the following. Suppose that
Alice, Bob, and Eve share a tripartite state $\rho_{\rABE}$, that Alice's
subsystem is limited to a two-dimensional Hilbert space, and that Alice
performs a Pauli $\sz$ measurement on her subsystem (in some chosen basis)
and flips the outcome with probability $q$. Then, the von Neumann entropy
$H(\sz|\rE)$ of Alice's outcome conditioned on Eve's quantum side information
is bounded by
\begin{IEEEeqnarray}{rCl}
  \label{eq:HZE-bb84-noise}
  H(\sz | \rE) &\geq& 1
  + \phi \Bigro{\sqrt{(1 - 2 q)^{2}
      + 4 q (1 - q) \abs{\avg{\sx \otimes B}}^{2}}} \nonumber \\
  &&-\> \phi \bigro{\abs{\avg{\sx \otimes B}}} \,,
\end{IEEEeqnarray}
where
\begin{equation}
  \avg{\sx \otimes B} = \Tr \bigsq{(\sx \otimes B) \rho_{\rAB}}
\end{equation}
is the correlation between the Pauli $\sx$ observable on Alice's side and any
$\pm 1$-valued observable $B$ on Bob's side computed on their part
$\rho_{\rAB}$ of the initial state $\rho_{\rABE}$. Note that, for $q = 0$,
\eqref{eq:HZE-bb84-noise} simplifies to the more familiar BB84 bound
\begin{equation}
  H(\sz | \rE) \geq 1 - \phi \bigro{\abs{\avg{\sx \otimes B}}}
\end{equation}
that we used in the outline in section~\ref{sec:simple-chsh}.

Before proving \eqref{eq:HZE-bb84-noise} we draw attention to a few of its
properties that are important for us here:
\begin{enumerate}
  \item \label{enum:nosym} \eqref{eq:HZE-bb84-noise} holds for any initial
  state $\rho_{\rABE}$. In particular, we do not assume that Alice's and
  Bob's marginal $\rho_{\rAB}$ must respect any symmetries or that the
  outcomes of any measurements they could perform on it must be equiprobable.
  \item \label{enum:monotonic} The right side of \eqref{eq:HZE-bb84-noise} is
  a monotonically increasing function in the argument
  $\abs{\avg{\sx \otimes B}}$. This means that if we know a (nonnegative)
  lower bound for the argument $\abs{\avg{\sx \otimes B}}$ then we can safely
  substitute it into \eqref{eq:HZE-bb84-noise} to obtain a lower bound for
  the conditional entropy.
  \item Although we will later only need to apply it to bipartite qubit
  systems, we remark that \eqref{eq:HZE-bb84-noise} is fully
  device-independent on Bob's side.
\end{enumerate}
A derivation of \eqref{eq:HZE-bb84-noise} written for the prepare-and-measure
version of the BB84 protocol that is device-independent on Bob's side already
exists \cite{ref:w2014}; we simply restate it here for the entanglement-based
setting that we are working in and modify it to confirm that the result still
holds even if Alice's measurement outcomes are not equiprobable, i.e., that
property~\ref{enum:nosym} holds. Property~\ref{enum:monotonic} only concerns
the end result and was already pointed out in \cite{ref:w2014}; appendix~B of
\cite{ref:w2014} in particular proves that \eqref{eq:HZE-bb84-noise} is
convex in the argument $\avg{\sx \otimes B}$ and attains its global minimum
at $\avg{\sx \otimes B} = 0$. This is also implied by
lemma~\ref{lem:convexity} in section~\ref{sec:convex-hull} of this article.

We start with the fact that we can assume Alice, Bob, and Eve initially share
a state $\ket{\Psi}_{\rABE}$ that is pure; this can be justified, for
instance, by the fact that the conditional entropy cannot increase if we
purify the initial state and give the extension to Eve. Next, using that
Alice's system is a qubit, we express the state as
\begin{equation}
  \label{eq:Psi_ABE}
  \ket{\Psi}_{\rABE} = \ket{0}_{\rA} \otimes \ket{\psi_{0}}_{\rBE}
  + \ket{1}_{\rA} \otimes \ket{\psi_{1}}_{\rBE} \,,
\end{equation}
where $\ket{0}$ and $\ket{1}$ are the eigenstates of $\sz$ and the states
$\ket{\psi_{0}}$ and $\ket{\psi_{1}}$ are subnormalised so that
$\norm{\psi_{0}}^{2} + \norm{\psi_{1}}^{2} = 1$. We don't assume
$\ket{\psi_{0}}$ and $\ket{\psi_{1}}$ are orthogonal to one another. The
correlation between Alice and Eve after Alice measures $\sz$ and flips the
outcome with probability $q$ is described by the classical-quantum state
\begin{equation}
  \label{eq:tau_ZE}
  \tau_{\rAE}
  = [0]_{\rA} \otimes (\bar{q} \psi^{\rE}_{0} + q \psi^{\rE}_{1})
  + [1]_{\rA} \otimes (q \psi^{\rE}_{0} + \bar{q} \psi^{\rE}_{1}) \,,
\end{equation}
where $\psi^{\rE}_{a} = \Tr_{\rB}[\psi_{a}]$.

To simplify the end result, we use that the conditional entropy
$H(\sz | \rE)_{\tau}$ of \eqref{eq:tau_ZE} is identical to the
conditional entropy $H(\sz | \rE)_{\tau'}$ of a state
\begin{equation}
  \tau'_{\rAE}
  = [1]_{\rA} \otimes (\bar{q} \psi^{\rE}_{0} + q \psi^{\rE}_{1})
  + [0]_{\rA} \otimes (q \psi^{\rE}_{0} + \bar{q} \psi^{\rE}_{1})
\end{equation}
which is identical to \eqref{eq:tau_ZE} except with $[0]$ and $[1]$
swapped. Furthermore, the entropy in both cases is the same as the
conditional entropy $H(\sz | \rE\rF)_{\bar{\tau}}$ computed on a symmetrised
state
\begin{equation}
  \label{eq:tau_ZEF}
  \bar{\tau}_{\rAE \rF}
  = \tfrac{1}{2} \tau_{\rAE} \otimes [0]_{\rF}
  + \tfrac{1}{2} \tau'_{\rAE} \otimes [1]_{\rF} \,.
\end{equation}
That is, one can verify that
\begin{equation}
  H(\sz | \rE\rF)_{\bar{\tau}}
  = \tfrac{1}{2} H(\sz | \rE)_{\tau} + \tfrac{1}{2} H(\sz | \rE)_{\tau'}
  = H(\sz | \rE)_{\tau} \,.
\end{equation}
Hence, we can bound $H(\sz | \rE)$ by deriving a lower bound for the
conditional entropy $H(\sz | \rE \rF)_{\bar{\tau}}$ of \eqref{eq:tau_ZEF}.

Grouping the terms in $[0]_{\rA}$ and $[1]_{\rA}$ together we rewrite
$\bar{\tau}$ as
\begin{equation}
  \label{eq:tau_ZEF_sigmas}
  \bar{\tau}_{\rAE\rF}
  = \tfrac{1}{2} \, [0]_{\rA}
  \otimes (\bar{q} \sigma_{=} + q \sigma_{\neq})
  + \tfrac{1}{2} \, [1]_{\rA}
  \otimes (q \sigma_{=} + \bar{q} \sigma_{\neq})
\end{equation}
with
\begin{IEEEeqnarray}{rCl}
  \sigma_{=} &=& \psi^{\rE}_{0} \otimes [0]_{\rF}
  + \psi^{\rE}_{1} \otimes [1]_{\rF} \,, \\
  \sigma_{\neq} &=& \psi^{\rE}_{1} \otimes [0]_{\rF}
  + \psi^{\rE}_{0} \otimes [1]_{\rF} \,,
\end{IEEEeqnarray}
which are normalised to $\Tr[\sigma_{=}] = \Tr[\sigma_{\neq}] = 1$. Next, we
use that
\begin{equation}
  H(\sz | \rE\rF) \geq H(\sz | \rBE\rF\rF')
\end{equation}
for any extension of \eqref{eq:tau_ZEF_sigmas}, i.e., any state
$\bar{\tau}_{\rABE\rF\rF'}$ such that
\begin{equation}
  \Tr_{\rB\rF'}[\bar{\tau}_{\rABE\rF\rF'}] = \bar{\tau}_{\rAE\rF} \,.
\end{equation}
Specifically, we use
\begin{equation}
  \label{eq:tau_ZBEFF}
  \bar{\tau}_{\rABE\rF\rF'}
  = \tfrac{1}{2} \, [0]_{\rA} \otimes (\bar{q} \chi_{=} + q \chi_{\neq})
  + \tfrac{1}{2} \, [1] _{\rA} \otimes (q \chi_{=} + \bar{q} \chi_{\neq})
\end{equation}
where we replace $\sigma_{=}$ and $\sigma_{\neq}$ in
\eqref{eq:tau_ZEF_sigmas} with purifications
\begin{IEEEeqnarray}{rCl}
  \ket{\chi_{=}} &=& \ket{\psi_{0}}_{\rBE} \otimes \ket{00}_{\rF\rF'}
  + \ket{\psi'_{1}}_{\rBE} \otimes \ket{11}_{\rF\rF'} \,, \\
  \ket{\chi_{\neq}} &=& \ket{\psi'_{1}}_{\rBE} \otimes \ket{00}_{\rF\rF'}
  + \ket{\psi_{0}}_{\rBE} \otimes \ket{11}_{\rF\rF'} \,,
  \IEEEeqnarraynumspace
\end{IEEEeqnarray}
where, in turn,
\begin{equation}
  \ket{\psi'_{1}} = B \otimes \id_{\rE} \ket{\psi_{1}}
\end{equation}
and $B$ is a (any) Hermitian operator satisfying $B^{2} = \id_{\rB}$. Direct
computation of the conditional entropy on the state \eqref{eq:tau_ZBEFF}
gives
\begin{IEEEeqnarray}{rCl}
  H(\sz | \rE) &\geq& H(\sz | \rB\rE\rF\rF') \nonumber \\
  &=& S(\bar{\tau}_{\rABE\rF\rF'})
  - S(\bar{\tau}_{\rBE\rF\rF'}) \nonumber \\
  &=& 1 + \tfrac{1}{2} S\bigro{\bar{q} \chi_{=} + q \chi_{\neq}}
  + \tfrac{1}{2} S\bigro{q \chi_{=} + \bar{q} \chi_{\neq}} \nonumber \\
  &&-\> S\bigro{\tfrac{1}{2}(\chi_{=} + \chi_{\neq})} \nonumber \\
  &=& 1 + \phi \Bigro{\sqrt{(\bar{q} - q)^{2}
      + 4 q \bar{q} \, \babs{\braket{\chi_{=}}{\chi_{\neq}}}^{2}}}
  \nonumber \\
  &&-\> \phi\bigro{\babs{\braket{\chi_{=}}{\chi_{\neq}}}} \,.
\end{IEEEeqnarray}
Finally, we obtain the result \eqref{eq:HZE-bb84-noise} by observing, using
the expression \eqref{eq:Psi_ABE} for the initial state $\ket{\Psi}_{\rABE}$,
that
\begin{IEEEeqnarray}{rCl}
  \braket{\chi_{=}}{\chi_{\neq}}
  &=& \bra{\psi_{0}} B \otimes \id_{\rE} \ket{\psi_{1}}
  + \bra{\psi_{1}} B \otimes \id_{\rE} \ket{\psi_{0}} \nonumber \\
  &=& \bra{\Psi} \sx \otimes B \otimes \id_{\rE} \ket{\Psi}_{\rABE}
  \nonumber \\
  &=& \avg{\sx \otimes B} \,.
\end{IEEEeqnarray}

Before returning to the device-independent protocol we remark that the BB84
entropy bound \eqref{eq:HZE-bb84-noise} is tight and can be attained with,
for example, $B = \sx$ and any tripartite state of the form
\begin{IEEEeqnarray}{rl}
  \label{eq:bb84-ZZ-XX-attack}
  \ket{\Psi}_{\rABE} = \frac{1}{2} \biggsq{
    &\sqrt{1 + E_{\rzz}} \sqrt{1 + E_{\rxx}} \,
    \ket{\phi_{+}}_{\rAB} \ket{++}_{\rE} \nonumber \\
    &+\> \sqrt{1 + E_{\rzz}} \sqrt{1 - E_{\rxx}} \,
    \ket{\phi_{-}}_{\rAB} \ket{+-}_{\rE} \nonumber \\
    &+\> \sqrt{1 - E_{\rzz}} \sqrt{1 + E_{\rxx}} \,
    \ket{\psi_{+}}_{\rAB} \ket{-+}_{\rE} \nonumber \\
    &+\> \sqrt{1 - E_{\rzz}} \sqrt{1 - E_{\rxx}} \,
    \ket{\psi_{-}}_{\rAB} \ket{--}_{\rE}} \,, \nonumber \\*
\end{IEEEeqnarray}
where
\begin{IEEEeqnarray}{rCl}
  \ket{\phi_{\pm}}
  &=& \frac{1}{\sqrt{2}} \Bigro{\ket{00} \pm \ket{11}} \,, \\
  \ket{\psi_{\pm}}
  &=& \frac{1}{\sqrt{2}} \Bigro{\ket{01} \pm \ket{10}}
\end{IEEEeqnarray}
are the Bell states, for $E_{\rxx} = \avg{\sx \otimes \sx}$ and any value
$-1 \leq E_{\rzz} \leq 1$ of $E_{\rzz} = \avg{\sz \otimes \sz}$. One can
verify that Alice's and Bob's marginal of \eqref{eq:bb84-ZZ-XX-attack} is
\begin{IEEEeqnarray}{rl}
  \label{eq:bb84-ZZ-XX-attack-AB}
  \Psi_{\rAB} = \frac{1}{4} \Bigsq{&
    \id \otimes \id + E_{\rxx} \, \sx \otimes \sx \nonumber \\
    &-\> E_{\rxx} E_{\rzz} \, \sy \otimes \sy
    + E_{\rzz} \, \sz \otimes \sz} \,.
\end{IEEEeqnarray}
This is the entanglement-based version of a family of optimal attacks
originally derived in the first security proof of the BB84 protocol against
individual attacks \cite{ref:fg1997}. The attack state
\eqref{eq:PsiABE-optattack} that we applied to the device-independent
protocol in section~\ref{sec:main-result} corresponds to the special case of
\eqref{eq:bb84-ZZ-XX-attack} with $E_{\rzz} = 1$. In both cases, the attack
strategy is independent of the amount of noise preprocessing applied.

\subsection{Correlations in the $\sz$-$\sx$ plane}
\label{sec:ZX-correlations}

As we saw in the outline, the BB84 bound effectively reduces the problem of
bounding the conditional entropy to applying quantum-mechanical constraints
on correlations that can appear in the subsystem shared by just Alice and
Bob. We show here that, for any underlying quantum state, the correlations
between the $\sz$ and $\sx$ Pauli operators always respect the bounds
\begin{IEEEeqnarray}{rCl}
  \label{eq:ZZ-ZX-constraint}
  E\du{\rzz}{2} + E\du{\rzx}{2} &\leq& 1 \,, \\
  \label{eq:XZ-XX-constraint}
  E\du{\rxz}{2} + E\du{\rxx}{2} &\leq& 1 \,,
\end{IEEEeqnarray}
and
\begin{IEEEeqnarray}{l}
  \label{eq:det-constraint}
  \bigro{1 - E\du{\rzz}{2} - E\du{\rzx}{2}}
  \bigro{1 - E\du{\rxz}{2} - E\du{\rxx}{2}}
  \nonumber \\
  \qquad \geq \bigro{E_{\rzz} E_{\rxz} + E_{\rzx} E_{\rxx}}^{2} \,,
\end{IEEEeqnarray}
where we use an abbreviated notation $E_{\rzz} = \avg{\sz \otimes \sz}$,
$E_{\rzx} = \avg{\sz \otimes \sx}$, and so on for the correlations. Note that
one of these constraints, \eqref{eq:ZZ-ZX-constraint}, is the constraint
\eqref{eq:ZZ-ZX-bound} that we used earlier in the outline.

To prove these constraints we use the fact that, for normalised Bloch
vectors $\vect{a} = (a_{\rz}, a_{\rx})$ and $\vect{b} = (b_{\rz}, b_{\rx})$,
the linear combinations $\vect{a} \cdot \sv$ and $\vect{b} \cdot \sv$ have
eigenvalues $\pm 1$. It follows that, for any state,
\begin{equation}
  \bavg{(\vect{a} \cdot \sv) \otimes (\vect{b} \cdot \sv)} \leq 1 \,.
\end{equation}
We can rewrite the left side as
\begin{IEEEeqnarray}{rCl}
  \bavg{(\vect{a} \cdot \sv) \otimes (\vect{b} \cdot \sv)}
  &=& \sum_{ij} a_{i} b_{j} \avg{\sigma_{i} \otimes \sigma_{j}} \nonumber \\
  &=& \mat{a}^{T} \mat{E} \, \mat{b} \,,
\end{IEEEeqnarray}
where $\mat{E}$ is the $2 \times 2$ matrix of coefficients
$E_{ij} = \avg{\sigma_{i} \otimes \sigma_{j}}$ for $i, j \in \{\rz,
\rx\}$. Since the relation
\begin{equation}
  \mat{a}^{T} \mat{E} \, \mat{b} \leq 1
\end{equation}
holds for any normalised vectors $\mat{a} = [a_{\rz}, a_{\rx}]^{T}$
and $\mat{b} = [b_{\rz}, b_{\rx}]^{T}$, it necessarily holds for whichever
vectors maximise the left side. Using these implies
\begin{equation}
  \norm{\mat{E}}_{\infty} \leq 1 \,.
\end{equation}
This is equivalent to the operator inequality $\mat{E} \mat{E}^{T} \leq \id$
or, put differently, that the matrix
\begin{equation}
  \id - \mat{E} \mat{E}^{T}
  = \begin{bmatrix}
    1 - E\du{\rzz}{2} - E\du{\rzx}{2} &
    - E_{\rzz} E_{\rxz} - E_{\rzx} E_{\rxx} \\
    - E_{\rzz} E_{\rxz} - E_{\rzx} E_{\rxx} &
    1 - E\du{\rxz}{2} - E\du{\rxx}{2}
  \end{bmatrix}
\end{equation}
is positive semidefinite. According to the Sylvester criterion, this is the
case if and only if all of its principal minors are of nonnegative
determinant, i.e., if
\begin{IEEEeqnarray}{rCl}
  1 - E\du{\rzz}{2} - E\du{\rzx}{2} &\geq& 0 \,, \\
  1 - E\du{\rxz}{2} - E\du{\rxx}{2} &\geq& 0 \,, \\
  \det \bigsq{\id - \mat{E} \mat{E}^{T}} &\geq& 0 \,.
\end{IEEEeqnarray}
These are exactly the constraints \eqref{eq:ZZ-ZX-constraint},
\eqref{eq:XZ-XX-constraint}, and \eqref{eq:det-constraint} asserted at the
beginning of this subsection.

\subsection{Entropy bound for qubits}
\label{sec:entropy-bound-qubits}

We are now ready to derive the bound satisfied by the conditional entropy for
qubit systems in terms of the $S_{\alpha}$ Bell expression. As we did in the
outline, we choose the bases of Alice's and Bob's systems such that their
measurement operators are of the form
\begin{IEEEeqnarray}{rCl}
  A_{1} &=& \sz \,, \\
  A_{2} &=& \cos(\phiA) \sz + \sin(\phiA) \sx
\end{IEEEeqnarray}
and
\begin{IEEEeqnarray}{rCl}
  B_{1} + B_{2} &=& 2 \cos\bigro{\tfrac{\phiB}{2}} \sz \,, \\
  B_{1} - B_{2} &=& 2 \sin\bigro{\tfrac{\phiB}{2}} \sx \,.
\end{IEEEeqnarray}
In this case the expectation value of $S_{\alpha}$ satisfies
\begin{IEEEeqnarray}{rCl}
  \label{eq:Sa-ZZ-ZX-XX-bound}
  S_{\alpha} / 2 &=& \cos \bigro{\tfrac{\phiB}{2}}
  \alpha \avg{\sz \otimes \sz}
  + \cos(\phiA) \sin \bigro{\tfrac{\phiB}{2}}
  \avg{\sz \otimes \sx}
  \nonumber \\
  &&+\> \sin(\phiA)
  \sin \bigro{\tfrac{\phiB}{2}} \avg{\sx \otimes \sx}
  \nonumber \\
  &\leq& \sqrt{\alpha^{2} \avg{\sz \otimes \sz}^{2}
    + \avg{\sz \otimes \sx}^{2} + \avg{\sx \otimes \sx}^{2}} \,.
  \IEEEeqnarraynumspace
\end{IEEEeqnarray}

For $\abs{\alpha} \geq 1$, the problem from this point is
straightforward. Using the constraint
\begin{IEEEeqnarray}{rCl}
  \avg{\sz \otimes \sz}^{2} + \avg{\sz \otimes \sx}^{2} \leq 1
\end{IEEEeqnarray}
from the previous subsection we obtain
\begin{equation}
  S\du{\alpha}{2}/4 \leq \alpha^{2} + \avg{\sx \otimes \sx}^{2} \,,
\end{equation}
which, making the choice $B = \sx$, rearranges to
\begin{equation}
  \abs{\avg{\sx \otimes B}} \geq
    \sqrt{S\du{\alpha}{2} / 4 - \alpha^{2}} \,.
\end{equation}
Using this in the BB84 entropy bound gives
\begin{IEEEeqnarray}{rCl}
  \IEEEeqnarraymulticol{3}{l}{
    \label{eq:HAE-qubits-alphage1}
    H(A_{1} | \rE)} \nonumber \\
  \quad &\geq& 1 + \phi \Bigro{\sqrt{(1 - 2q)^{2}
      + 4 q (1 - q) (S\du{\alpha}{2}/4 - \alpha^{2})}} \nonumber \\
  &&-\> \phi \Bigro{\sqrt{S\du{\alpha}{2}/4 - \alpha^{2}}}
\end{IEEEeqnarray}
for all $\abs{\alpha} \geq 1$ for qubits, and we only need to verify that the
right side is convex in $S_{\alpha}$ to justify extending the result to
arbitrary dimension, which we do in the next subsection.

For $\abs{\alpha} < 1$ we need to do a bit more work. In this case, we choose
$B$ to be of the form $\cos(\theta) \sz + \sin(\theta) \sx$ such that
\begin{equation}
  \avg{\sx \otimes B} = \cos(\theta) \avg{\sx \otimes \sz}
  + \sin(\theta) \avg{\sx \otimes \sx} \,.
\end{equation}
For the best $\theta$,
\begin{equation}
  \abs{\avg{\sx \otimes B}}
  = \sqrt{\avg{\sx \otimes \sz}^{2} + \avg{\sx \otimes \sx}^{2}} \,.
\end{equation}
Together with \eqref{eq:Sa-ZZ-ZX-XX-bound}, and using the notation and
constraints derived in the previous section, the full problem we want to
solve is
\begin{equation}
  \label{eq:avgXB-optimisation}
  \begin{IEEEeqnarraybox}[][c]{rCul}
    E_{\alpha} (S_{\alpha}) &=&
    min. & \sqrt{E\du{\rxz}{2} + E\du{\rxx}{2}} \\
    && s.t. & \left\{
      \begin{IEEEeqnarraybox}[][c]{rCl}
        \alpha^{2} E\du{\rzz}{2} + E\du{\rzx}{2} + E\du{\rxx}{2}
        &\geq& S\du{\alpha}{2}/4 \\
        E\du{\rzz}{2} + E\du{\rzx}{2} &\leq& 1 \\
        E\du{\rxz}{2} + E\du{\rxx}{2} &\leq& 1 \\
        \begin{IEEEeqnarraybox}[][b]{l}
          (1 - E\du{\rzz}{2} - E\du{\rzx}{2}) \\
          \times\> (1 - E\du{\rxz}{2} - E\du{\rxx}{2})
        \end{IEEEeqnarraybox}
        && \\
        -\> (E_{\rzz} E_{\rxz} + E_{\rzx} E_{\rxx})^{2} &\geq& 0
      \end{IEEEeqnarraybox}
    \right.
  \end{IEEEeqnarraybox}
\end{equation}
in the variables $E_{\rzz}$, $E_{\rzx}$, $E_{\rxz}$, $E_{\rxx}$. The solution
to this optimisation problem is derived in detail in
appendix~\ref{sec:solution}. The end result, depending on $S_{\alpha}$, is
\begin{equation}
  \label{eq:qubit-f-alpha-defge}
  E_{\alpha}(S_{\alpha}) = \sqrt{S\du{\alpha}{2} / 4 - \alpha^{2}}
\end{equation}
for $\abs{S_{\alpha}} \geq 2 \sqrt{1 + \alpha^{2} - \alpha^{4}}$ and
\begin{equation}
  \label{eq:qubit-f-alpha-defle}
  E_{\alpha}(S_{\alpha})
  = \sqrt{1 - \Bigro{1 - \tfrac{1}{\abs{\alpha}} \sqrt{(1 - \alpha^{2})
      (S\du{\alpha}{2}/4 - 1)}}^{2}}
\end{equation}
for $\abs{S_{\alpha}} \leq 2 \sqrt{1 + \alpha^{2} - \alpha^{4}}$. Applying
this in the BB84 bound \eqref{eq:HZE-bb84-noise} gives
\begin{IEEEeqnarray}{rCl}
  \label{eq:HAE-qubit-bound}
  H(A_{1} | \rE)
  &\geq& 1
  + \phi \Bigro{\sqrt{(1 - 2q)^{2} + 4q(1 - q) E_{\alpha}(S_{\alpha})^{2}}}
  \nonumber \\
  &&-\> \phi \bigro{E_{\alpha}(S_{\alpha})} \,,
\end{IEEEeqnarray}
with $E_{\alpha}(S_{\alpha})$ given by \eqref{eq:qubit-f-alpha-defge} or
\eqref{eq:qubit-f-alpha-defle} depending on the value of $S_{\alpha}$.

As a side remark, we note that the lower bounds on
$\abs{\avg{\sx \otimes B}} $ that we just derived in term of $S_{\alpha}$ can
be used to derive the tight bound for the min-entropy in terms of
$S_{\alpha}$. This is discussed in appendix~\ref{sec:min-entropy}.

\subsection{Device-independent entropy bound}
\label{sec:convex-hull}

Having bounded the conditional entropy for qubit systems, the remaining step
is to establish the convexity with respect to the Bell expectation value
$S_{\alpha}$, or to construct the convex hull, of the family of bounds we
have derived.

\begin{figure}[htbp]
  \centering
  \includegraphics{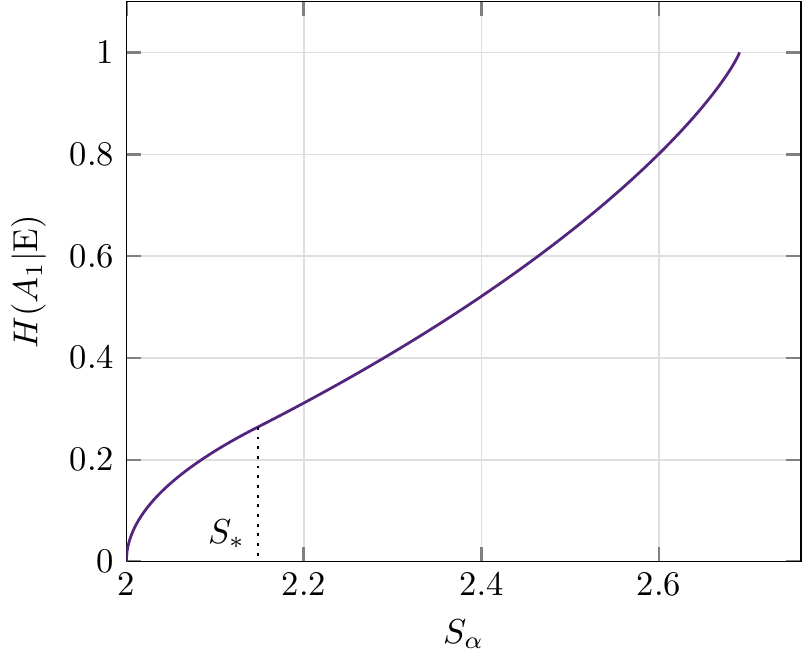}
  \caption{Conditional entropy bound \eqref{eq:HAE-qubit-bound} for $q = 0$
    and $\alpha = 0.9$ derived assuming Alice and Bob perform projective
    qubit measurements. Its form depends on how $S_{\alpha}$ compares to
    $S_{*} = 2 \sqrt{1 + \alpha^{2} - \alpha^{4}} \approx 2.1484$: it is
    concave and described by \eqref{eq:HAE-qubit-bound} and
    \eqref{eq:qubit-f-alpha-defle} for $S_{\alpha} \leq S_{*}$ and it is
    convex and described by \eqref{eq:HAE-qubit-bound} and
    \eqref{eq:qubit-f-alpha-defge} for $S_{\alpha} \geq S_{*}$.}
  \label{fig:HA_E-Salpha-qubits}
\end{figure}

The qubit bound \eqref{eq:HAE-qubit-bound} is illustrated for $q = 0$ and
$\alpha = 0.9$ in figure~\ref{fig:HA_E-Salpha-qubits}. It visibly has the
appearance of being concave for
$S_{\alpha} \leq 2 \sqrt{1 + \alpha^{2} - \alpha^{4}}$ and convex for
$S_{\alpha}$ above this value. We show here that this is generally true of
\eqref{eq:HAE-qubit-bound} for all $q$ and all $\abs{\alpha} < 1$ while the
qubit bound \eqref{eq:HAE-qubits-alphage1} for $\abs{\alpha} \geq 1$, which
has the same form as \eqref{eq:HAE-qubit-bound} for $\abs{\alpha} < 1$ and
$S_{\alpha} \geq 2 \sqrt{1 + \alpha^{2} - \alpha^{4}}$, is always convex.

We establish the concavity or convexity of the qubit bounds by bounding
their second derivatives. To do this, we recall some conditions under which
concavity or convexity are preserved under function composition. The second
derivative of the composition $f \circ g$ of two functions is given by
\begin{equation}
  (f \circ g)''(x) = f'' \bigro{g(x)} \, g'(x)^{2}
  + f' \bigro{g(x)} \, g''(x) \,.
\end{equation}
From this we can see that $f \circ g$ is guaranteed to be convex if both $f$
and $g$ are convex and if $f$ is monotonically increasing. Conversely,
$f \circ g$ is guaranteed to be concave if $f$ is concave and monotonically
decreasing while $g$ is convex.

Using this approach, we prove that the bound given by
\eqref{eq:HAE-qubit-bound} and \eqref{eq:qubit-f-alpha-defge} as well as
\eqref{eq:HAE-qubits-alphage1} is convex by expressing it as
$f \circ g(S_{\alpha})$ for the function $f$ in lemma~\ref{lem:convexity}
below with $Q = (1 - 2q)^{2}$ and with
\begin{equation}
  g(S_{\alpha}) = S\du{\alpha}{2}/4 - \alpha^{2} \,,
\end{equation}
which is clearly convex\footnote{Note that using the BB84 bound for $f$ and
  using $g(S_{\alpha}) = \sqrt{S\du{\alpha}{2}/4 - \alpha^{2}}$ for $g$ does
  not work with this approach since $g$ would be concave in that case.}. The
following result, which we prove in appendix~\ref{sec:convexity}, confirms
that $f$ has the properties needed for us to infer that the composition
$f \circ g$ is convex.
\begin{lemma}
  \label{lem:convexity}
  The function
  \begin{equation}
    \label{eq:f-convexity}
    f(x) = 1 + \phi \Bigro{\sqrt{Q + (1 - Q) x}}
    - \phi \bigro{\sqrt{x}}
  \end{equation}
  is convex and monotonically increasing in $x$ for $0 \leq x \leq 1$ and for
  any $0 \leq Q \leq 1$.
\end{lemma}

We similarly prove that the curve described by \eqref{eq:HAE-qubit-bound} and
\eqref{eq:qubit-f-alpha-defle} is concave by expressing it as
$f \circ g(S_{\alpha})$, this time with the function $f$ in
lemma~\ref{lem:concavity} below with $Q = (1 - 2q)^{2}$ and with
\begin{equation}
  \label{eq:g_concavity}
  g(S_{\alpha}) = 1 - \frac{1}{\abs{\alpha}}
  \sqrt{(1 - \alpha^{2}) (S\du{\alpha}{2}/4 - 1)} \,.
\end{equation}
Checking that $g$ is convex amounts to checking that the function
$s \mapsto \sqrt{s^{2} - 1}$ is concave, which doesn't present any particular
problem. The following result, proved in appendix~\ref{sec:concavity},
verifies that $f$ has the properties required to guarantee that $f \circ g$
is concave.
\begin{lemma}
  \label{lem:concavity}
  The function
  \begin{equation}
    \label{eq:f-concavity}
    f(x) = 1 + \phi\Bigro{\sqrt{Q + (1 - Q) (1 - x^{2})}}
    - \phi\Bigro{\sqrt{1 - x^{2}}}
  \end{equation}
  is concave and monotonically decreasing in $x$ for $0 \leq x \leq 1$ and
  for any $0 \leq Q \leq 1$.
\end{lemma}

Finally, one can verify that the qubit entropy bound for $\abs{\alpha} < 1$
and its first derivative in $S_{\alpha}$ are continuous. This amounts to
checking that \eqref{eq:qubit-f-alpha-defge} and
\eqref{eq:qubit-f-alpha-defle} both have the same values and first
derivatives,
\begin{IEEEeqnarray}{rCl}
  E_{\alpha}(S_{*}) &=& \sqrt{1 - \alpha^{4}} \,, \\
  E'_{\alpha}(S_{*})
  &=& \frac{\sqrt{1 + \alpha^{2} - \alpha^{4}}}{2 \sqrt{1 - \alpha^{4}}} \,,
\end{IEEEeqnarray}
at the point $S_{*} = 2 \sqrt{1 + \alpha^{2} - \alpha^{4}}$. One can also
verify that its gradient becomes infinite as $S_{\alpha}$ approaches the
quantum bound $2 \sqrt{1 + \alpha^{2}}$.

The device-independent bound in arbitrary dimension is given by the convex
hull of the qubit bound. This implies that the part of the qubit bound
described by \eqref{eq:HAE-qubit-bound} and \eqref{eq:qubit-f-alpha-defle}
for $S_{\alpha} \leq 2 \sqrt{1 + \alpha^{2} - \alpha^{4}}$, which is concave,
may be ignored and the device-independent bound is thus given by the
construction described at the end of section~\ref{sec:main-result} and
illustrated in figure~\ref{fig:HA_E-Salpha}. In particular, this is where the
lower limit of $2 \sqrt{1 + \alpha^{2} - \alpha^{4}}$ in the range
\eqref{eq:s*-root-range} for the root-finding problem
\eqref{eq:s*-root-problem} comes from. The fact that the qubit entropy bound
and its gradient are continuous everywhere, that it is concave for
$S_{\alpha} \leq 2 \sqrt{1 + \alpha^{2} - \alpha^{4}}$ and reaches $h(q)$ at
the classical bound $S_{\alpha} = 2$, that it is convex for
$S_{\alpha} \geq 2 \sqrt{1 + \alpha^{2} - \alpha^{4}}$, and that its gradient
becomes infinite at the quantum bound imply that there is necessarily a
solution to the root-finding problem \eqref{eq:s*-root-problem} in the range
\eqref{eq:s*-root-range} and that it is unique.

\section{Applications to DIQKD key rates}
\label{sec:keyrate-examples}

The entropy bound $H(A_{1} | \rE) \geq \bar{g}_{q,\alpha}(S_{\alpha})$ we
have now proved can be applied in QKD security frameworks that reduce proving
the security of a protocol to bounding the conditional von Neumann entropy in
a single round. Applying it in the Devetak-Winter rate
\eqref{eq:devetak-winter} gives a lower bound
\begin{equation}
  \label{eq:dw-Salpha-bound}
  r \geq \bar{g}_{q,\alpha}(S_{\alpha}) - H(A_{1} | B_{3})
\end{equation}
on the asymptotic key rate that depends only on parameters -- the Bell
expectation value $S_{\alpha}$ and probabilities $P(ab|13)$ -- that Alice and
Bob working together can estimate.

In this section, we apply \eqref{eq:dw-Salpha-bound} to obtain explicit
estimates of the robustness of the device-independent QKD protocol in two
commonly studied imperfection models, both of which were also used as
examples in \cite{ref:pa2009}: depolarising noise, where we assume that the
optimal Bell state for the protocol is mixed with white noise, and a generic
loss model.

All the thresholds we report when using noise preprocessing were computed in
the limit $q \to 1/2$ of maximal random noise. This typically seems to give
the best threshold and this was what we saw in cases where we computed the
key rate for different amounts of noise preprocessing, although we have not
checked that $q \to 1/2$ is optimal in every case. We describe how the
Devetak-Winter rate can be computed in this limit in
appendix~\ref{sec:max-preprocessing}.

\subsection{Depolarising noise}


\begin{figure}[tbp]
  \centering
  \includegraphics{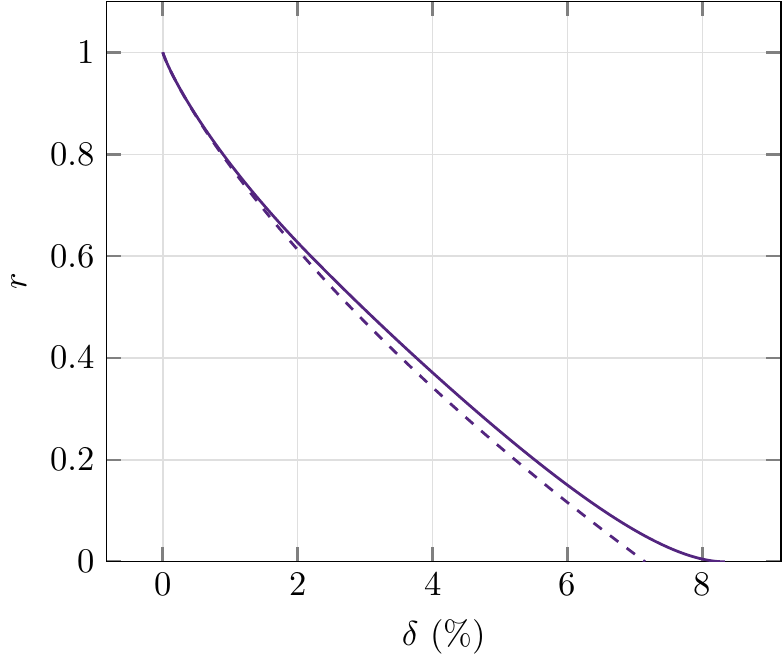}
  \caption{Lower bound \eqref{eq:dw-depolarising-noise} on the Devetak-Winter
    rate as a function of channel error rate $\delta$, assuming correlations
    satisfying
    $\avg{A_{1} (B_{1} + B_{2})} = \avg{A_{1} (B_{1} - B_{2})} = \sqrt{2} (1
    - 2\delta)$, for the optimal values of $q$ and $\alpha$ (solid curve) and
    for $q = 0$ and $\alpha = 1$ (dashed curve).}
  \label{fig:dw-error}
\end{figure}

In this model we suppose that Alice and Bob share a noisy version,
\begin{equation}
  \label{eq:rho-visibility}
  \rho_{\rAB} = v \, \phi_{+} + (1 - v) \, \id_{\rAB}/4 \,,
\end{equation}
of the optimal two-qubit Bell state
\begin{equation}
  \ket{\phi_{+}} = \frac{1}{\sqrt{2}} \bigro{\ket{00} + \ket{11}}
\end{equation}
parametrised by some visibility $v$. For the ideal measurements $A_{1} = \sz$
and $B_{3} = \sz$ for key generation, the possible outcomes are obtained with
joint probabilities
\begin{IEEEeqnarray}{rCcCl}
  \label{eq:Pab13-delta}
  P({+}{+} | 1 3) &=& P({-}{-} | 1 3) &=& (1 - \delta) / 2 \,, \\
  P({+}{-} | 1 3) &=& P({-}{+} | 1 3) &=& \delta / 2 \,,
\end{IEEEeqnarray}
where the error rate $\delta$ is related to the visibility in
\eqref{eq:rho-visibility} by
\begin{equation}
  v = 1 - 2 \delta \,.
\end{equation}
When Alice additionally applies noise preprocessing, the resulting joint
distribution retains the same form but with a worse error rate,
\begin{equation}
  \delta_{q} = q + (1 - 2 q) \delta \,.
\end{equation}
The conditional Shannon entropy associated with this distribution is
\begin{equation}
  H(A_{1} | B_{3}) = h(\delta_{q}) \,,
\end{equation}
depending on the amount $q$ of noise preprocessing applied.

In the CHSH-based protocol, the ideal measurements in the Bell test are
$A_{1} = \sz$, $A_{2} = \sx$, and $B_{1,2} = (\sz \pm \sx) / \sqrt{2}$. With
these measurements the two-body correlation terms satisfy
\begin{equation}
  \bavg{A_{1} (B_{1} + B_{2})} = \bavg{A_{1} (B_{1} - B_{2})} =
  \sqrt{2} (1 - 2 \delta) \,,
\end{equation}
which translates to an expectation value
\begin{equation}
  \label{eq:Salpha-whitenoise}
  S_{\alpha} = \sqrt{2} (1 + \alpha) (1 - 2 \delta)
\end{equation}
of the asymmetric CHSH expression.

The lower bound on the Devetak-Winter rate we obtain for the
depolarising noise model is then explicitly
\begin{equation}
  \label{eq:dw-depolarising-noise}
  r \geq \bar{g}_{q,\alpha} \bigro{\sqrt{2} (1 + \alpha) (1 - 2 \delta)}
  - h(\delta_{q}) \,.
\end{equation}
The best possible bound on the key rate is obtained by maximising the right
side of \eqref{eq:dw-depolarising-noise} over $\alpha$ and $q$. We illustrate
the result as a function of the channel noise rate $\delta$ in
figure~\ref{fig:dw-error}. The key rate computed using only the CHSH bound of
\cite{ref:pa2009}, i.e., $q = 0$ and $\alpha = 1$, is also shown for
comparison. The combination of applying noise preprocessing and optimising
over the $S_{\alpha}$ family of Bell expressions increases the threshold
error rate, up to which the key rate remains positive, from
$\delta \approx 7.15\%$ found in \cite{ref:pa2009} to $8.33\%$.

\begin{table}[btp]
  \centering
  \begin{tabular}{l|cc}
                            & $q = 0$ & $q \to 1/2$ \\ \hline
    $\alpha = 1$            &  7.1492 &      8.0848 \\
    $\alpha = \opt$         &  7.4002 &      8.3320 \\
    $\alpha, B_{y} = \opt$  &  7.4177 &      8.3453 \\
  \end{tabular}
  \caption{Threshold error rates (\%) obtained using either CHSH
    ($\alpha = 1$) or the optimal asymmetric expression ($\alpha = \opt$),
    both without ($q = 0$) and with maximal ($q \to 1/2$) noise
    preprocessing. The third row ($\alpha, B_{y} = \opt$) gives the
    thresholds when in addition Bob's measurements are optimised such that
    $S_{\alpha} = 2 \sqrt{1 + \alpha^{2}} (1 - 2 \delta)$.}
  \label{tab:noise-thresholds}
\end{table}

In table~\ref{tab:noise-thresholds} we list the threshold error rates
obtained for the different combinations of using CHSH or the optimal
$S_{\alpha}$ expressions without or with noise
preprocessing. Table~\ref{tab:noise-thresholds} in addition gives the
thresholds obtained when using, instead of the measurements
$B_{1,2} = (\sz \pm \sx) / \sqrt{2}$ that are optimal for CHSH, the
measurements that attain the maximal value
\begin{equation}
  S_{\alpha} = 2 \sqrt{1 + \alpha^{2}} (1 - 2 \delta)
\end{equation}
of the $S_{\alpha}$ expression for the depolarised state. This gives
marginally better threshold error rates.

Since the conditional entropy bounds used in the above security analysis are
tight, the threshold error rates that we compute are optimal in terms of the
asymmetric CHSH expressions $S_\alpha$, and the values reported in
table~\ref{tab:noise-thresholds} optimised over $\alpha$ are optimal in terms
of the combinations $\avg{A_{1} B_{1}} + \avg{A_{1} B_{2}}$ and
$\avg{A_{2} B_{1}} - \avg{A_{2} B_{2}}$ viewed as independent parameters. But
they are actually also optimal with respect to an analysis that would take
into account the full set of statistics. This is because according to the
measurement and noise model considered above, Alice's and Bob's marginal
measurement outcomes are equiprobable, i.e.,
\begin{equation}\label{eq:optim-marg}
  \avg{A_{1}} = \avg{A_{2}} = \avg{B_{1}} = \avg{B_{2}} = 0 \,,
\end{equation}
and the two-body correlations satisfy
\begin{equation}\label{eq:optim-a1b}
  \avg{A_{1} B_{1}} = \avg{A_{1} B_{2}}
\end{equation}
and
\begin{equation}\label{eq:optim-a2b}
  \avg{A_{2} B_{1}} = - \avg{A_{2} B_{2}} \,.
\end{equation}
But these relations, which completely fix the full set of correlators once
the independent combinations $\avg{A_{1} B_{1}} + \avg{A_{1} B_{2}}$ and
$\avg{A_{2} B_{1}} - \avg{A_{2} B_{2}}$ are specified, are also satisfied for
the family of optimal attacks presented in section~\ref{sec:main-result} and
saturating our entropy bound. Thus, specifying other correlation terms beyond
those involved in the definition of $S_{\alpha}$ would not restrict the
attack strategies further than already considered.

\subsection{Losses}

%
%
%

In this setting, we suppose that Alice and Bob detect their particles and
obtain definite measurement outcomes with some probability $\eta$ which, for
simplicity, we take to be the same on both sides. We model this formally by
treating nondetection events as a third measurement outcome, obtained
independently by Alice and Bob with probability $1 - \eta$. In this case, as
well as the maximally-entangled Bell state we also consider a possible type
of strategy in which Alice and Bob deliberately use partially-entangled
states, which have been shown to improve the robustness of Bell experiments
based on the CHSH inequality to losses \cite{ref:e1993}.

We consider the maximally-entangled state first. In order to apply our
entropy bound we need to reduce the setting to one where the measurements
used in the Bell test all have only two outcomes. The typical way to do this,
which we apply here, is to map (``bin'') nondetection events to one of the
outcomes $+1$ or $-1$. In terms of the global detection
efficiency $\eta$, the maximum value of the $S_{\alpha}$ expression over the
different possible binning strategies is
\begin{equation}
  S_{\alpha} = \sqrt{2} (1 + \alpha) \eta^{2}
  + 2 \max(1, \abs{\alpha}) \bar{\eta}^{2} \,,
\end{equation}
where $\bar{\eta} = 1 - \eta$, if Bob uses the diagonal measurements
$B_{1,2} = (\sz \pm \sx) / \sqrt{2}$ or
\begin{equation}
  S_{\alpha} = 2 \sqrt{1 + \alpha^{2}} \, \eta^{2}
  + 2 \max(1, \abs{\alpha}) \bar{\eta}^{2}
\end{equation}
if Bob uses the optimal ones. For the key generation measurements $A_{1} =
B_{3} = \sz$, Alice and Bob obtain outcomes (including nondetections) with
the joint probabilities
\begin{equation}
  \label{eq:PAB-losses}
  \bigro{P_{\rAB}(ab | 13)} = \begin{bmatrix}
    \tfrac{1}{2} \eta^{2} & 0 & \tfrac{1}{2} \eta \bar{\eta} \\
    0 & \tfrac{1}{2} \eta^{2} & \tfrac{1}{2} \eta \bar{\eta} \\
    \tfrac{1}{2} \eta \bar{\eta} & \tfrac{1}{2} \eta \bar{\eta}
    & \bar{\eta}^{2}
  \end{bmatrix} \,;
\end{equation}
however, since we map nondetection events to (for example) $A_{1} = +1$ on
Alice's side to use the entropy bound we must do the same here, that is, we
should add the third row of \eqref{eq:PAB-losses} to the first. This gives
the joint distribution
\begin{equation}
  \label{eq:PAB-losses-binned}
  \bigro{P_{\rAB}(ab | 13)} =
  \begin{bmatrix}
    \tfrac{1}{2} \eta & \tfrac{1}{2} \eta \bar{\eta}
    & \tfrac{1}{2} \bar{\eta} (1 + \bar{\eta}) \\
    0 & \tfrac{1}{2} \eta^{2} & \tfrac{1}{2} \eta \bar{\eta}
  \end{bmatrix} \,.
\end{equation}
Finally, as before, when noise preprocessing is applied we also need to swap
the rows of \eqref{eq:PAB-losses-binned} with probability $q$, i.e.,
transform \eqref{eq:PAB-losses-binned} according to
\begin{equation}
  P_{\rAB}(\pm, b | 13) \mapsto (1 - q) P_{\rAB}(\pm, b | 13)
  + q P_{\rAB}(\mp, b | 13) \,,
\end{equation}
before computing the conditional Shannon entropy $H(A_{1} | B_{3})$.

\begin{table}[tbp]
  \centering
  \begin{tabular}{l|cc}
                             & $q = 0$ & $q \to 1/2$ \\ \hline
    $\alpha = 1$             & 90.7768 &     90.3046 \\
    $\alpha = \opt$          & 90.4970 &     90.0230 \\
    $\alpha, B_{y} = \opt$   & 90.4856 &     90.0122 \\
  \end{tabular}
  \caption{Threshold detection efficiencies (\%) obtained both without
    ($q = 0$) and with maximal ($q \to 1/2$) noise preprocessing for the
    maximally-entangled state. The first ($\alpha = 1$) and second
    ($\alpha = \opt$) rows give the thresholds obtained using only CHSH and
    the optimal asymmetric Bell expression using diagonal measurements
    $(\sz \pm \sx)/\sqrt{2}$ on Bob's side. In the third row
    ($\alpha, B_{y} = \opt$) we also use the optimal measurements on Bob's
    side.}
  \label{tab:eta-singlet}
\end{table}

The threshold global detection efficiencies we found for the resulting
Devetak-Winter rate for the maximally-entangled state are reported in
table~\ref{tab:eta-singlet}. In this case the thresholds are all a little
over $90\%$ with little variation depending on whether the $S_{\alpha}$
family or noise preprocessing are used. The threshold $\eta \approx 90.78\%$
that we obtain using only CHSH and with no noise preprocessing is better than
the threshold $\eta \approx 92.4\%$ found in \cite{ref:pa2009} as a result of
computing the conditional Shannon entropy on the full probability
distribution \eqref{eq:PAB-losses-binned} without binning the nondetection
event on Bob's side. It is also slightly better than the threshold of
$90.9\%$ found in \cite{ref:ml2012} due to a small advantage in bounding the
Devetak-Winter rate via the conditional von Neumann entropy rather than via
the Holevo quantity as was originally done in \cite{ref:pa2009}.

We now consider partially-entangled states which, as we mentioned, are known
to increase the robustness to losses in the CHSH Bell experiment. In this
case, we suppose that Alice and Bob share a state
\begin{equation}
  \label{eq:psi-theta}
  \ket{\psi_{\theta}} = \cos \bigro{\tfrac{\theta}{2}} \ket{00}
  + \sin \bigro{\tfrac{\theta}{2}} \ket{11}
\end{equation}
dependent on a parameter $\theta$ characterising the degree of
entanglement. The density operator associated to $\ket{\psi_{\theta}}$ is
\begin{IEEEeqnarray}{rl}
  \psi_{\theta} = \frac{1}{4} \biggsq{&
    \id \otimes \id + \cos(\theta) \bigro{\sz \otimes \id + \id \otimes \sz}
    \nonumber \\
    &+\> \sin(\theta) \bigro{\sx \otimes \sx - \sy \otimes \sy}
    + \sz \otimes \sz} \,. \IEEEeqnarraynumspace
\end{IEEEeqnarray}
We then suppose that Alice and Bob measure $A_{1} = \sz$ and $B_{3} = \sz$ to
generate their key and use whichever measurements $A_{2}$, $B_{1}$, and
$B_{2}$ give the highest expectation value of the $S_{\alpha}$ expression
given that $A_{1}$ is fixed to $\sz$ and the global detection efficiency is
fixed to some value $\eta$. For this problem, the best thresholds we saw were
obtained by mapping all nondetection events to $+1$. For this binning
strategy, the expectation value of $S_{\alpha}$ can be expressed as
\begin{IEEEeqnarray}{rCl}
  S_{\alpha} &=& \eta^{2} \bavg{
    \alpha A_{1} (B_{1} + B_{2}) + A_{2} (B_{1} - B_{2})} \nonumber \\
  &&+\> \eta \bar{\eta} \bavg{2\alpha A_{1}
    + (\alpha + 1) B_{1} + (\alpha - 1) B_{2}} \nonumber \\
  &&+\> 2 \bar{\eta}^{2} \alpha
\end{IEEEeqnarray}
in terms of $\eta$ and the expectation values $\avg{A_{x}}$, $\avg{B_{y}}$,
and $\avg{A_{x} B_{y}}$ that would be obtained from \eqref{eq:psi-theta} if
there were no losses. Setting
\begin{equation}
  A_{2} = \cos(\phiA) \sz + \sin(\phiA) \sx
\end{equation}
and optimising over the measurements $B_{1}$ and $B_{2}$ on Bob's side gives
\begin{IEEEeqnarray}{rCl}
  \label{eq:Salpha-psitheta}
  S_{\alpha} &=& \eta \sqrt{R^{2} + (P + Q)^{2}}
  + \eta \sqrt{R^{2} + (P - Q)^{2}} \nonumber \\
  &&+\> 2 \eta \bar{\eta} \alpha \cos(\theta)
  + 2 \bar{\eta}^{2} \alpha \,,
\end{IEEEeqnarray}
where
\begin{IEEEeqnarray}{rCl}
    R &=& \eta \sin(\phiA) \sin(\theta) \,, \\
    P &=& \alpha \eta + \alpha \bar{\eta} \cos(\theta) \,, \\
    Q &=& \eta \cos(\phiA) + \bar{\eta} \cos(\theta) \,,
\end{IEEEeqnarray}
in terms of $\theta$, $\phiA$, and $\eta$. With this strategy, for small
$\theta$\footnote{More precisely, the approximation
  \eqref{eq:chsh-small-theta} is valid if $\abs{\theta}$ is small compared to
  $\abs{\phiA}$. This means that $\phiA$ can be taken arbitrarily close to
  zero as long as $\theta$ is taken even smaller. This condition is also why
  \eqref{eq:chsh-small-theta} does not imply that the CHSH inequality can be
  violated with $\phiA = 0$.} the expectation value in the special case of
CHSH is approximated by
\begin{equation}
  \label{eq:chsh-small-theta}
  S \approx 2 + \eta \biggsq{
    3 \eta - 2
    - \frac{\eta \bar{\eta} \bigro{1 - \cos(\phiA)}}
      {2 - \eta \bigro{1 - \cos(\phiA)}}} \theta^{2}
\end{equation}
to the smallest nontrivial order in $\theta$, or
\begin{equation}
  \label{eq:chsh-small-angles}
  S \approx 2 + \eta \Bigro{
    3 \eta - 2 - \frac{1}{4} \eta \bar{\eta} {\phiA}^{2}}
  \theta^{2}
\end{equation}
if $\phiA$ is also small. This shows that the strategy we have described can
violate the CHSH inequality as long as the global detection efficiency is
better than $\eta = 2/3$, the same as was found in \cite{ref:e1993}, although
our choice to fix $A_{1} = \sz$ means that the CHSH violation we can attain
is not as high as it could otherwise be.

The outcomes including nondetections when Alice and Bob measure $A_{1} = \sz$
and $B_{3} = \sz$ on the partially-entangled state occur with joint
probabilities
\begin{equation}
  \bigro{P_{\rAB}(ab | 13)} = \begin{bmatrix}
    \eta^{2} c^{2} & 0 & \eta \bar{\eta} c^{2} \\
    0 & \eta^{2} s^{2} & \eta \bar{\eta} s^{2} \\
    \eta \bar{\eta} c^{2} & \eta \bar{\eta} s^{2} & \bar{\eta}^{2}
  \end{bmatrix}
\end{equation}
where $c^{2} = \cos\bigro{\tfrac{\theta}{2}}^{2}$ and
$s^{2} = \sin\bigro{\tfrac{\theta}{2}}^{2}$. As before, we should merge the
nondetection events on Alice's side with the $+1$ outcome and swap the rows
with probability $q$ if noise preprocessing is also used before computing
$H(A_{1} | B_{3})$.

Computing the Devetak-Winter rate using the value \eqref{eq:Salpha-psitheta}
of $S_{\alpha}$ and maximising the result over $\theta$ and $\phiA$ gives a
positive rate up to the global detection efficiencies listed in
table~\ref{tab:eta-psitheta}. The thresholds for $q = 0$ are attained for
partially-entangled states with $\theta$ a little under $0.5$ radians. The
threshold for $q \to 1/2$ by contrast is attained in the limit $\theta \to 0$
of a separable state. The approximations of the key rate for
$q = (1 - \varepsilon)/2$ described in appendix~\ref{sec:max-preprocessing}
and \eqref{eq:chsh-small-angles} of CHSH for small $\theta$ and $\phiA$ can
be used to derive an approximate lower bound,
\begin{equation}
  r \gtrsim \frac{\eta}{6 \log(2)} \Bigro{
    3 \eta^{2} + 6 \eta - 7
    - \frac{1}{2} \eta \bar{\eta} {\phiA}^{2}}
  \theta^{2} \varepsilon^{2} \,,
\end{equation}
for the key rate when $\varepsilon$ and the angles are small. In this
vicinity the key rate can be positive, albeit minuscule, as long as the
global detection efficiency is better than
\begin{equation}
  \eta = \sqrt{10/3} - 1 \approx 82.5742\% \,.
\end{equation}
For $q \to 1/2$ we didn't see any improvement to the threshold when using the
$S_{\alpha}$ family instead of the CHSH expression.

\begin{table}[tbp]
  \centering
  \begin{tabular}{l|cc}
                     & $q = 0$ & $q \to 1/2$ \\ \hline
    $\alpha = 1$     & 86.5479 &   82.5742 \\
    $\alpha = \opt$  & 86.5255 &   82.5742 \\
  \end{tabular}
  \caption{Threshold detection efficiencies (\%) obtained using either CHSH
    ($\alpha = 1$) or the optimal asymmetric expression ($\alpha = \opt$),
    both without ($q = 0$) and with maximal ($q \to 1/2$) noise
    preprocessing, for the strategy using partially-entangled states.}
  \label{tab:eta-psitheta}
\end{table}

\begin{figure}[htbp]
  \centering
  \includegraphics{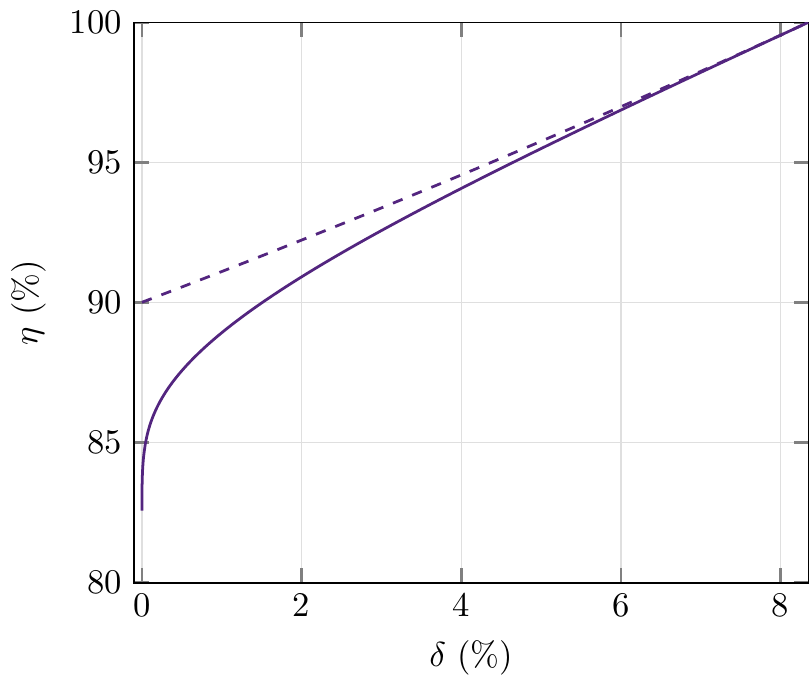}
  \caption{Threshold detection efficiency as a function of channel noise when
    partially-entangled states (solid curve) or maximally-entangled states
    (dashed curve) are used. The thresholds start respectively at
    $\eta \approx 82.5742\%$ and $\eta \approx 90.0122\%$ for $\delta = 0$
    and increase to $\eta = 100\%$ as the error rate approaches
    $\delta \approx 8.3453\%$. $\delta$ is defined here to be related to the
    visibility in \eqref{eq:rho-theta-visibility} by $v = 1 - 2 \delta$.}
  \label{fig:critical-eta}
\end{figure}

The results in table~\ref{tab:eta-psitheta} should be taken with a pinch of
salt as they were derived assuming only losses occur
in an otherwise perfect experiment, which is not realistic. The threshold
detection efficiency using noise preprocessing in particular was derived by
taking the limit $\theta \to 0$ of a separable state and is accordingly very
vulnerable to noise. To model this, we computed the best thresholds (i.e.,
using both noise preprocessing and the $S_{\alpha}$ family) when we replace
the initial state with an attenuated one of the form
\begin{equation}
  \label{eq:rho-theta-visibility}
  \rho = v \, \psi_{\theta} + (1 - v) \, \id_{\rAB}/4 \,.
\end{equation}
The threshold detection efficiencies both for $\theta = \pi/2$ (the
maximally-entangled state) and for whichever partially-entangled state gave
the best result are illustrated as a function of the error rate in
figure~\ref{fig:critical-eta}. The threshold using partially-entangled states
visibly increases very rapidly as soon as we add even a small amount of
channel noise. We also recomputed the thresholds of
table~\ref{tab:eta-psitheta} with the visibility set to $v = 99\%$,
corresponding to a more realistic error rate of $\delta = 0.5\%$. This
increases the thresholds, listed in table~\ref{tab:eta-psitheta-noise}, to
above $87\%$.

Finally, note that while the conditional entropy bound we used holds
generally, it is only really optimised for the case that Alice and Bob's
correlations satisfy Eqs.~\eqref{eq:optim-marg}--\eqref{eq:optim-a2b} and in
particular obtain equiprobable measurement outcomes. Deterministically
binning nondetection events and deliberately using a partially-entangled
state both spoil this and the real thresholds could actually be significantly
better than the ones we report here.

\begin{table}[ht]
  \centering
  \begin{tabular}{l|cc}
                     & $q = 0$ & $q \to 1/2$ \\ \hline
    $\alpha = 1$     & 88.8316 &     87.6469 \\
    $\alpha = \opt$  & 88.7149 &     87.5714 \\
  \end{tabular}
  \caption{Threshold detection efficiencies (\%) obtained using either CHSH
    ($\alpha = 1$) or the optimal asymmetric expression ($\alpha = \opt$),
    both without ($q = 0$) and with maximal ($q \to 1/2$) noise
    preprocessing, using partially-entangled states but with a $0.5\%$
    channel error rate.}
  \label{tab:eta-psitheta-noise}
\end{table}

\section{Discussion}

In our work we derived a tight lower bound on the conditional von Neumann
entropy following an arbitrary amount of noise preprocessing and for the
family $S_{\alpha}$ of asymmetric CHSH Bell expressions, which allows us to
make more effective use of the statistics than when using the standard CHSH
expression. Our proof heavily exploited the similarity of the
device-independent protocol to the entanglement-based version of the BB84
protocol. Section~\ref{sec:keyrate-examples} showed that these modifications,
both individually and together, can improve the robustness of the original
CHSH-based protocol using two commonly-used imperfection models as
examples. For a maximally entangled two-qubit state subject to a depolarising
noise model, the threshold error rate according to our analysis is just above
$8.34\%$. This is actually the optimal error rate, equalling a security
analysis that takes into account the full set of statistics.

As is typically the case of research based on the CHSH Bell setting, our
analysis is heavily dependent on the fact that the setting can be effectively
reduced to the study of bipartite qubit systems. Obviously, it would be
interesting in the future to learn how to derive good bounds for the
conditional von Neumann entropy in Bell settings with more inputs and/or
outputs, where we cannot rely on such a reduction.

Within the CHSH setting however there are still some possible avenues for
further work. First, while the entropy bound we have derived is tight in
terms of the parameters it depends on, this does not mean it is always
optimal for every scenario. Our approach in particular is optimised for the
case that Alice's and Bob's marginal measurement outcomes are
equiprobable. This is fine if the imperfections in a real implementation most
closely correspond to the depolarising noise model but not, as we cautioned
in section~\ref{sec:keyrate-examples}, if they more closely resemble the loss
model. It is likely that our entropy bound gives suboptimal results in the
latter case. This is partly confirmed by a recent result for
partially-entangled qubits in \cite{ref:bff2021}, where a threshold of about
$84.3\%$ is obtained for the global detection efficiency without using noise
preprocessing, which is somewhat better than the thresholds of around
$86.5\%$ that we obtained for this case here.

Our proof, however, has a rather modular nature; parts of it could no doubt
be changed, generalised, or applied to different problems without affecting
other parts. Different preprocessings could be considered and may only
require changing the derivation of the BB84 bound in
section~\ref{sec:bb84-noise-bound}; we have not checked, for instance, if
flipping both of Alice's outcomes with the same probability $q$ is always the
optimal choice. Optimisation problems of the kind we landed on in
section~\ref{sec:entropy-bound-qubits} may lend themselves to numerical
approaches\footnote{In particular, Eq.~\eqref{eq:avgXB-optimisation} as
  written is the square root of a polynomial optimisation problem and could
  in principle be solved numerically using the Lasserre hierarchy
  \cite{ref:l2001}. This would still be true, albeit the problem larger, if
  we had not optimised out the measurements; the sines and cosines of the
  angles $\phiA$ and $\phiB$ could still be treated as additional variables
  satisfying polynomial constraints $c^{2} + s^{2} = 1$.}, although it should
be kept in mind that solving the problem analytically made it much more
straightforward for us to prove when the result was and was not
convex. Lemmas~\ref{lem:convexity} and \ref{lem:concavity} in
section~\ref{sec:convex-hull} may help prove the convexity or nonconvexity of
entropy bounds with similar functional forms to what we derived in
section~\ref{sec:derivation}.

Second, our approach exploits two refinements -- using more information about
the statistics and noise preprocessing -- that were already known to improve
the performance of cryptography protocols. A third refinement that we have
not exploited here would consist of using both of Alice's measurements to
generate the key, which forces an eavesdropper to have to gain information
about both bases without knowing in advance which will be used. This kind of
modification has previously been shown to improve the average bound on the
min-entropy in the device-independent setting \cite{ref:bss2014}.

This variant of the CHSH-based protocol has recently been considered
\cite{ref:sg2020}, however the approach of \cite{ref:sg2020} requires a
rather elaborate numerical procedure to bound the key rate and the threshold
error rate of $8.2\%$ reported by the authors for the depolarising channel
does not exceed the threshold just above $8.34\%$ that we found for the
single-basis version of the protocol using the refinements we considered
here.

We suspect that the result of \cite{ref:sg2020} is not quite optimal,
however, and thus a good candidate for further study. One possible way to
bound the average entropy of Alice's measurements for this problem may be to
try to apply the same method we have applied to the single-basis protocol
here. According to a quick numerical test we performed, the best bound on the
average conditional entropy that could be obtained using only the BB84
entropy bound of section~\ref{sec:bb84-noise-bound} and Pauli correlation
bounds of section~\ref{sec:ZX-correlations} should give a slightly better
threshold of around $8.36\%$, or alternatively up to $9.24\%$ if noise
preprocessing is also used. Even these thresholds do not appear to be
optimal, however. We also performed a brute-force numerical minimisation of
the average conditional von Neumann entropy. The results seemed to show that
the optimal attack for qubit systems involves Alice and Bob using
measurements of the form
$A_{1,2} = \cos \bigro{\tfrac{\phiA}{2}} \sz \pm \sin
\bigro{\tfrac{\phiA}{2}} \sx$ and $B_{1,2} = \sz, \sx$ on an
asymmetric version of the optimal BB84 attack state\footnote{Section~I.H of
  the supplementary information to Ref.~\cite{ref:sg2020} conjectures that
  the reduced state shared by Alice and Bob in the optimal attack is Bell
  diagonal with two nonzero eigenvalues, which would correspond to an attack
  state like \eqref{eq:bb84-ZZ-XX-attack} with, e.g., $E_{\rzz} = 1$, but
  this is not consistent with what we found when minimising the average
  conditional von Neumann entropy directly. The minimum of
  \eqref{eq:HA_E-avg-2bases} subject to \eqref{eq:CHSH-2bases} is generally
  not attained with either $E_{\rzz} = \pm 1$ or $E_{\rxx} = \pm 1$.}, i.e.,
\eqref{eq:bb84-ZZ-XX-attack} with different values of $E_{\rzz}$ and
$E_{\rxx}$. In other words, the tight lower bound on the average entropy for
qubit systems appeared to us to coincide with the result of minimising
\begin{IEEEeqnarray}{rCl}
  \IEEEeqnarraymulticol{3}{l}{
    \label{eq:HA_E-avg-2bases}
    \tfrac{1}{2} H(A_{1} | \rE) + \tfrac{1}{2} H(A_{2} | \rE)} \nonumber \\
  \qquad &=& 1 + \phi \Bigro{\sqrt{
      \cos \bigro{\tfrac{\phiA}{2}}^{2} E\du{\rzz}{2}
      + \sin \bigro{\tfrac{\phiA}{2}}^{2} E\du{\rxx}{2}}} \nonumber \\
  &&-\> \phi(E_{\rzz}) - \phi(E_{\rxx})
\end{IEEEeqnarray}
subject to
\begin{equation}
  \label{eq:CHSH-2bases}
  2 \cos \bigro{\tfrac{\phiA}{2}} E_{\rzz}
  + 2 \sin \bigro{\tfrac{\phiA}{2}} E_{\rxx} = S
\end{equation}
for a given expectation value $S$ of the CHSH correlator. Furthermore,
similar to the qubit bound we derived in
section~\ref{sec:entropy-bound-qubits}, the resulting bound appears to be
concave except in a region close to the quantum bound, where the optimal
qubit attack appears to consist of mutually unbiased measurements
($\phiA = \pi/2$) on a state for which $E_{\rzz} = E_{\rxx} =
S/\sqrt{8}$. Assuming these observations are correct, this would mean that
the average bound for qubit systems is given by
\begin{equation}
  \label{eq:HA_E-avg-conjecture}
  \tfrac{1}{2} H(A_{1} | \rE) + \tfrac{1}{2} H(A_{2} | \rE)
  \geq 1 - \phi \bigro{S/\sqrt{8}}
\end{equation}
if the CHSH expectation value is close to the quantum maximum, with the
device-independent bound obtained by extending one of the tangents of
\eqref{eq:HA_E-avg-conjecture} as in the construction of our main result in
section~\ref{sec:main-result}. This result would imply that the threshold
noise rate of the DIQKD protocol using both bases (without noise
preprocessing) is around $8.44\%$.

Eqs.~\eqref{eq:HA_E-avg-2bases} and \eqref{eq:CHSH-2bases} suggest it may be
difficult to rigorously prove the tight bound on the average conditional
entropy for the two-basis version of the DIQKD protocol. Nevertheless, the
thresholds we have estimated numerically suggest there is some room for
improvement in the results of \cite{ref:sg2020}, particularly if noise
preprocessing is also used.

\paragraph{Note added.} A derivation of the conditional entropy bound for
CHSH incorporating noise preprocessing, i.e., the special case $\alpha = 1$
of the conditional entropy bound we derive here for the full $S_{\alpha}$
family, has recently been published in \cite{ref:hs2020} independently of us,
which the authors apply to an investigation of the performance of an optical
model. This entropy bound is obtained by parametrising and explicitly
optimising over all qubit attacks, following the approach of
\cite{ref:ab2007,ref:pa2009}. Here, we exploited the fact that we already
know how to derive the entropy bound including noise preprocessing for the
BB84 protocol \cite{ref:w2014,ref:w2014b}. The qubit analysis of
\cite{ref:hs2020} can be promoted to a fully, dimension-free,
device-independent bound using the convexity proof we give in
appendix~\ref{sec:concavexity} (as this step is incomplete in
\cite{ref:hs2020}).

After making public the present results, a follow-up to \cite{ref:hs2020}
providing an analytical derivation of the same entropic bounds for the
$S_{\alpha}$ expressions for $\alpha \geq 1$ and proposing a numerical method
for $\abs{\alpha} < 1$ has appeared in~\cite{ref:sb2020}.

Finally, the conjecture described around \eqref{eq:HA_E-avg-conjecture} on
the average entropy of both of Alice's measurement outcomes was also made
independently in \cite{ref:brc2021}.

\section*{Acknowledgements}

This work was supported by the Spanish MINECO (Severo Ochoa grant
SEV-2015-0522 and TRANQI PID2019-106888GB-I00 / AEI / 10.13039/501100011033),
the Generalitat de Catalunya (CERCA Program, QuantumCAT and SGR 1381),
Fundacio Privada Cellex and Mir-Puig, the AXA Chair in Quantum Information
Science, the ERC AdG CERQUTE, and the EU Quantum Flagship project
QRANGE\@. S.P. is a Senior Research Associate of the Fonds de la Recherche
Scientifique -- FNRS\@.

\bibliography{references}

\begin{thebibliography}{37}%
\makeatletter
\providecommand \@ifxundefined [1]{%
 \@ifx{#1\undefined}
}%
\providecommand \@ifnum [1]{%
 \ifnum #1\expandafter \@firstoftwo
 \else \expandafter \@secondoftwo
 \fi
}%
\providecommand \@ifx [1]{%
 \ifx #1\expandafter \@firstoftwo
 \else \expandafter \@secondoftwo
 \fi
}%
\providecommand \natexlab [1]{#1}%
\providecommand \enquote  [1]{``#1''}%
\providecommand \bibnamefont  [1]{#1}%
\providecommand \bibfnamefont [1]{#1}%
\providecommand \citenamefont [1]{#1}%
\providecommand \href@noop [0]{\@secondoftwo}%
\providecommand \href [0]{\begingroup \@sanitize@url \@href}%
\providecommand \@href[1]{\@@startlink{#1}\@@href}%
\providecommand \@@href[1]{\endgroup#1\@@endlink}%
\providecommand \@sanitize@url [0]{\catcode `\\12\catcode `\$12\catcode
  `\&12\catcode `\#12\catcode `\^12\catcode `\_12\catcode `\%12\relax}%
\providecommand \@@startlink[1]{}%
\providecommand \@@endlink[0]{}%
\providecommand \url  [0]{\begingroup\@sanitize@url \@url }%
\providecommand \@url [1]{\endgroup\@href {#1}{\urlprefix }}%
\providecommand \urlprefix  [0]{URL }%
\providecommand \Eprint [0]{\href }%
\providecommand \doibase [0]{https://doi.org/}%
\providecommand \selectlanguage [0]{\@gobble}%
\providecommand \bibinfo  [0]{\@secondoftwo}%
\providecommand \bibfield  [0]{\@secondoftwo}%
\providecommand \translation [1]{[#1]}%
\providecommand \BibitemOpen [0]{}%
\providecommand \bibitemStop [0]{}%
\providecommand \bibitemNoStop [0]{.\EOS\space}%
\providecommand \EOS [0]{\spacefactor3000\relax}%
\providecommand \BibitemShut  [1]{\csname bibitem#1\endcsname}%
\let\auto@bib@innerbib\@empty
\bibitem [{\citenamefont {Bell}(1964)}]{ref:b1964}%
  \BibitemOpen
  \bibfield  {author} {\bibinfo {author} {\bibfnamefont {J.~S.} \bibnamefont
  {Bell}}, }\href {http://cds.cern.ch/record/111654/} {\bibfield  {journal}
  {\bibinfo  {journal} {\physics} }\textbf {\bibinfo {volume} {1}}, \bibinfo
  {pages} {195} (\bibinfo {year} {1964})}\BibitemShut {NoStop}%
\bibitem [{\citenamefont {Brunner} \emph {et~al.}(2014)\citenamefont {Brunner},
  \citenamefont {Cavalcanti}, \citenamefont {Pironio}, \citenamefont {Scarani},
  and \citenamefont {Wehner}}]{ref:bc2014}%
  \BibitemOpen
  \bibfield  {author} {\bibinfo {author} {\bibfnamefont {N.}~\bibnamefont
  {Brunner}}, \bibinfo {author} {\bibfnamefont {D.}~\bibnamefont {Cavalcanti}},
  \bibinfo {author} {\bibfnamefont {S.}~\bibnamefont {Pironio}}, \bibinfo
  {author} {\bibfnamefont {V.}~\bibnamefont {Scarani}},  and \bibinfo {author}
  {\bibfnamefont {S.}~\bibnamefont {Wehner}}, }\href
  {\doibase10.1103/RevModPhys.86.419} {\bibfield  {journal} {\bibinfo
  {journal} {\rmp} }\textbf {\bibinfo {volume} {86}}, \bibinfo {pages} {419}
  (\bibinfo {year} {2014})}, \Eprint {http://arxiv.org/abs/1303.2849}
  {ar{X}iv:1303.2849 [quant-ph]}\BibitemShut {NoStop}%
\bibitem [{\citenamefont {Ac\'\i{}n} \emph {et~al.}(2007)\citenamefont
  {Ac\'\i{}n}, \citenamefont {Brunner}, \citenamefont {Gisin}, \citenamefont
  {Massar}, \citenamefont {Pironio}, and \citenamefont {Scarani}}]{ref:ab2007}%
  \BibitemOpen
  \bibfield  {author} {\bibinfo {author} {\bibfnamefont {A.}~\bibnamefont
  {Ac\'\i{}n}}, \bibinfo {author} {\bibfnamefont {N.}~\bibnamefont {Brunner}},
  \bibinfo {author} {\bibfnamefont {N.}~\bibnamefont {Gisin}}, \bibinfo
  {author} {\bibfnamefont {S.}~\bibnamefont {Massar}}, \bibinfo {author}
  {\bibfnamefont {S.}~\bibnamefont {Pironio}},  and \bibinfo {author}
  {\bibfnamefont {V.}~\bibnamefont {Scarani}}, }\href
  {\doibase10.1103/PhysRevLett.98.230501} {\bibfield  {journal} {\bibinfo
  {journal} {\prl} }\textbf {\bibinfo {volume} {98}}, \bibinfo {pages} {230501}
  (\bibinfo {year} {2007})}, \Eprint {http://arxiv.org/abs/quant-ph/0702152}
  {ar{X}iv:quant-ph/0702152}\BibitemShut {NoStop}%
\bibitem [{\citenamefont {Pironio} \emph {et~al.}(2009)\citenamefont {Pironio},
  \citenamefont {Ac\'\i{}n}, \citenamefont {Brunner}, \citenamefont {Gisin},
  \citenamefont {Massar}, and \citenamefont {Scarani}}]{ref:pa2009}%
  \BibitemOpen
  \bibfield  {author} {\bibinfo {author} {\bibfnamefont {S.}~\bibnamefont
  {Pironio}}, \bibinfo {author} {\bibfnamefont {A.}~\bibnamefont {Ac\'\i{}n}},
  \bibinfo {author} {\bibfnamefont {N.}~\bibnamefont {Brunner}}, \bibinfo
  {author} {\bibfnamefont {N.}~\bibnamefont {Gisin}}, \bibinfo {author}
  {\bibfnamefont {S.}~\bibnamefont {Massar}},  and \bibinfo {author}
  {\bibfnamefont {V.}~\bibnamefont {Scarani}}, }\href
  {\doibase10.1088/1367-2630/11/4/045021} {\bibfield  {journal} {\bibinfo
  {journal} {\njp} }\textbf {\bibinfo {volume} {11}}, \bibinfo {pages} {045021}
  (\bibinfo {year} {2009})}, \Eprint {http://arxiv.org/abs/0903.4460}
  {ar{X}iv:0903.4460 [quant-ph]}\BibitemShut {NoStop}%
\bibitem [{\citenamefont {Ekert}(1991)}]{ref:e1991}%
  \BibitemOpen
  \bibfield  {author} {\bibinfo {author} {\bibfnamefont {A.~K.} \bibnamefont
  {Ekert}}, }\href {\doibase10.1103/PhysRevLett.67.661} {\bibfield  {journal}
  {\bibinfo  {journal} {\prl} }\textbf {\bibinfo {volume} {67}}, \bibinfo
  {pages} {661} (\bibinfo {year} {1991})}\BibitemShut {NoStop}%
\bibitem [{\citenamefont {Clauser} \emph {et~al.}(1969)\citenamefont {Clauser},
  \citenamefont {Horne}, \citenamefont {Shimony}, and \citenamefont
  {Holt}}]{ref:ch1969}%
  \BibitemOpen
  \bibfield  {author} {\bibinfo {author} {\bibfnamefont {J.~F.} \bibnamefont
  {Clauser}}, \bibinfo {author} {\bibfnamefont {M.~A.} \bibnamefont {Horne}},
  \bibinfo {author} {\bibfnamefont {A.}~\bibnamefont {Shimony}},  and \bibinfo
  {author} {\bibfnamefont {R.~A.} \bibnamefont {Holt}}, }\href
  {\doibase10.1103/PhysRevLett.23.880} {\bibfield  {journal} {\bibinfo
  {journal} {\prl} }\textbf {\bibinfo {volume} {23}}, \bibinfo {pages} {880}
  (\bibinfo {year} {1969})}\BibitemShut {NoStop}%
\bibitem [{\citenamefont {Arnon-Friedman} \emph {et~al.}(2018)\citenamefont
  {Arnon-Friedman}, \citenamefont {Dupuis}, \citenamefont {Fawzi},
  \citenamefont {Renner}, and \citenamefont {Vidick}}]{ref:afd2018}%
  \BibitemOpen
  \bibfield  {author} {\bibinfo {author} {\bibfnamefont {R.}~\bibnamefont
  {Arnon-Friedman}}, \bibinfo {author} {\bibfnamefont {F.}~\bibnamefont
  {Dupuis}}, \bibinfo {author} {\bibfnamefont {O.}~\bibnamefont {Fawzi}},
  \bibinfo {author} {\bibfnamefont {R.}~\bibnamefont {Renner}},  and \bibinfo
  {author} {\bibfnamefont {T.}~\bibnamefont {Vidick}}, }\href
  {\doibase10.1038/s41467-017-02307-4} {\bibfield  {journal} {\bibinfo
  {journal} {\natcomm} }\textbf {\bibinfo {volume} {9}}, \bibinfo {pages} {459}
  (\bibinfo {year} {2018})}\BibitemShut {NoStop}%
\bibitem [{\citenamefont {Zhang} \emph {et~al.}(2020)\citenamefont {Zhang},
  \citenamefont {Fu}, and \citenamefont {Knill}}]{ref:zfk2020}%
  \BibitemOpen
  \bibfield  {author} {\bibinfo {author} {\bibfnamefont {Y.}~\bibnamefont
  {Zhang}}, \bibinfo {author} {\bibfnamefont {H.}~\bibnamefont {Fu}},  and
  \bibinfo {author} {\bibfnamefont {E.}~\bibnamefont {Knill}}, }\href
  {\doibase10.1103/PhysRevResearch.2.013016} {\bibfield  {journal} {\bibinfo
  {journal} {\prresearch} }\textbf {\bibinfo {volume} {2}}, \bibinfo {pages}
  {013016} (\bibinfo {year} {2020})}, \Eprint {http://arxiv.org/abs/1806.04553}
  {ar{X}iv:1806.04553 [quant-ph]}\BibitemShut {NoStop}%
\bibitem [{\citenamefont {Devetak} and \citenamefont
  {Winter}(2005)}]{ref:dw2005}%
  \BibitemOpen
  \bibfield  {author} {\bibinfo {author} {\bibfnamefont {I.}~\bibnamefont
  {Devetak}} and \bibinfo {author} {\bibfnamefont {A.}~\bibnamefont {Winter}},
  }\href {\doibase10.1098/rspa.2004.1372} {\bibfield  {journal} {\bibinfo
  {journal} {\prsa} }\textbf {\bibinfo {volume} {461}}, \bibinfo {pages} {207}
  (\bibinfo {year} {2005})}, \Eprint {http://arxiv.org/abs/quant-ph/0306078}
  {ar{X}iv:quant-ph/0306078}\BibitemShut {NoStop}%
\bibitem [{\citenamefont {Renner} \emph {et~al.}(2005)\citenamefont {Renner},
  \citenamefont {Gisin}, and \citenamefont {Kraus}}]{ref:rgk2005}%
  \BibitemOpen
  \bibfield  {author} {\bibinfo {author} {\bibfnamefont {R.}~\bibnamefont
  {Renner}}, \bibinfo {author} {\bibfnamefont {N.}~\bibnamefont {Gisin}},  and
  \bibinfo {author} {\bibfnamefont {B.}~\bibnamefont {Kraus}}, }\href
  {\doibase10.1103/PhysRevA.72.012332} {\bibfield  {journal} {\bibinfo
  {journal} {\pra} }\textbf {\bibinfo {volume} {72}}, \bibinfo {pages} {012332}
  (\bibinfo {year} {2005})}, \Eprint {http://arxiv.org/abs/quant-ph/0502064}
  {ar{X}iv:quant-ph/0502064}\BibitemShut {NoStop}%
\bibitem [{\citenamefont {Tan} \emph {et~al.}(2019)\citenamefont {Tan},
  \citenamefont {Schwonnek}, \citenamefont {Goh}, \citenamefont {Primaatmaja},
  and \citenamefont {Lim}}]{ref:ts2019}%
  \BibitemOpen
  \bibfield  {author} {\bibinfo {author} {\bibfnamefont {E.~Y.-Z.} \bibnamefont
  {Tan}}, \bibinfo {author} {\bibfnamefont {R.}~\bibnamefont {Schwonnek}},
  \bibinfo {author} {\bibfnamefont {K.~T.} \bibnamefont {Goh}}, \bibinfo
  {author} {\bibfnamefont {I.~W.} \bibnamefont {Primaatmaja}},  and \bibinfo
  {author} {\bibfnamefont {C.~C.-W.} \bibnamefont {Lim}}, }\href@noop {} {
  (\bibinfo {year} {2019})}, \Eprint {http://arxiv.org/abs/1908.11372}
  {arXiv:1908.11372 [quant-ph]}\BibitemShut {NoStop}%
\bibitem [{\citenamefont {Schwonnek} \emph {et~al.}(2020)\citenamefont
  {Schwonnek}, \citenamefont {Goh}, \citenamefont {Primaatmaja}, \citenamefont
  {Tan}, \citenamefont {Wolf}, \citenamefont {Scarani}, and \citenamefont
  {Lim}}]{ref:sg2020}%
  \BibitemOpen
  \bibfield  {author} {\bibinfo {author} {\bibfnamefont {R.}~\bibnamefont
  {Schwonnek}}, \bibinfo {author} {\bibfnamefont {K.~T.} \bibnamefont {Goh}},
  \bibinfo {author} {\bibfnamefont {I.~W.} \bibnamefont {Primaatmaja}},
  \bibinfo {author} {\bibfnamefont {E.~Y.-Z.} \bibnamefont {Tan}}, \bibinfo
  {author} {\bibfnamefont {R.}~\bibnamefont {Wolf}}, \bibinfo {author}
  {\bibfnamefont {V.}~\bibnamefont {Scarani}},  and \bibinfo {author}
  {\bibfnamefont {C.~C.-W.} \bibnamefont {Lim}}, }\href@noop {} {  (\bibinfo
  {year} {2020})}, \Eprint {http://arxiv.org/abs/2005.02691}
  {ar{X}iv:2005.02691 [quant-ph]}\BibitemShut {NoStop}%
\bibitem [{\citenamefont {Grasselli} \emph {et~al.}(2021)\citenamefont
  {Grasselli}, \citenamefont {Murta}, \citenamefont {Kampermann}, and
  \citenamefont {Bru\ss{}}}]{ref:gm2021}%
  \BibitemOpen
  \bibfield  {author} {\bibinfo {author} {\bibfnamefont {F.}~\bibnamefont
  {Grasselli}}, \bibinfo {author} {\bibfnamefont {G.}~\bibnamefont {Murta}},
  \bibinfo {author} {\bibfnamefont {H.}~\bibnamefont {Kampermann}},  and
  \bibinfo {author} {\bibfnamefont {D.}~\bibnamefont {Bru\ss{}}}, }\href
  {\doibase10.1103/PRXQuantum.2.010308} {\bibfield  {journal} {\bibinfo
  {journal} {\prxquantum} }\textbf {\bibinfo {volume} {2}}, \bibinfo {pages}
  {010308} (\bibinfo {year} {2021})}, \Eprint {http://arxiv.org/abs/2004.14263}
  {ar{X}iv:2004.14263 [quant-ph]}\BibitemShut {NoStop}%
\bibitem [{\citenamefont {Brown} \emph {et~al.}(2021)\citenamefont {Brown},
  \citenamefont {Fawzi}, and \citenamefont {Fawzi}}]{ref:bff2021}%
  \BibitemOpen
  \bibfield  {author} {\bibinfo {author} {\bibfnamefont {P.}~\bibnamefont
  {Brown}}, \bibinfo {author} {\bibfnamefont {H.}~\bibnamefont {Fawzi}},  and
  \bibinfo {author} {\bibfnamefont {O.}~\bibnamefont {Fawzi}}, }\href
  {\doibase10.1038/s41467-020-20018-1} {\bibfield  {journal} {\bibinfo
  {journal} {\natcomm} }\textbf {\bibinfo {volume} {12}}, \bibinfo {pages}
  {575} (\bibinfo {year} {2021})}, \Eprint {http://arxiv.org/abs/2007.12575}
  {ar{X}iv:2007.12575 [quant-ph]}\BibitemShut {NoStop}%
\bibitem [{\citenamefont {Bennett} and \citenamefont
  {Brassard}(1984)}]{ref:bb1984}%
  \BibitemOpen
  \bibfield  {author} {\bibinfo {author} {\bibfnamefont {C.~H.} \bibnamefont
  {Bennett}} and \bibinfo {author} {\bibfnamefont {G.}~\bibnamefont
  {Brassard}}, }in \href@noop {} {\emph {\bibinfo {booktitle} {Proceedings of
  IEEE International Conference on Computers, Systems and Signal Processing}}}
  (\bibinfo  {publisher} {IEEE}, \bibinfo {address} {New York}, \bibinfo {year}
  {1984}) pp. \bibinfo {pages} {175--179}, \Eprint
  {http://arxiv.org/abs/2003.06557} {ar{X}iv:2003.06557 [quant-ph]}\BibitemShut
  {NoStop}%
\bibitem [{\citenamefont {Murta} \emph {et~al.}(2019)\citenamefont {Murta},
  \citenamefont {van Dam}, \citenamefont {Ribeiro}, \citenamefont {Hanson}, and
  \citenamefont {Wehner}}]{ref:mvd2019}%
  \BibitemOpen
  \bibfield  {author} {\bibinfo {author} {\bibfnamefont {G.}~\bibnamefont
  {Murta}}, \bibinfo {author} {\bibfnamefont {S.~B.} \bibnamefont {van Dam}},
  \bibinfo {author} {\bibfnamefont {J.}~\bibnamefont {Ribeiro}}, \bibinfo
  {author} {\bibfnamefont {R.}~\bibnamefont {Hanson}},  and \bibinfo {author}
  {\bibfnamefont {S.}~\bibnamefont {Wehner}}, }\href
  {\doibase10.1088/2058-9565/ab2819} {\bibfield  {journal} {\bibinfo  {journal}
  {\qscitech} }\textbf {\bibinfo {volume} {4}}, \bibinfo {pages} {035011}
  (\bibinfo {year} {2019})}, \Eprint {http://arxiv.org/abs/1811.07983}
  {ar{X}iv:1811.07983 [quant-ph]}\BibitemShut {NoStop}%
\bibitem [{\citenamefont {Ko\l{}ody\'{n}ski} \emph {et~al.}(2020)\citenamefont
  {Ko\l{}ody\'{n}ski}, \citenamefont {M\'{a}ttar}, \citenamefont {Skrzypczyk},
  \citenamefont {Woodhead}, \citenamefont {Cavalcanti}, \citenamefont
  {Banaszek}, and \citenamefont {Ac\'{\i}n}}]{ref:km2020}%
  \BibitemOpen
  \bibfield  {author} {\bibinfo {author} {\bibfnamefont {J.}~\bibnamefont
  {Ko\l{}ody\'{n}ski}}, \bibinfo {author} {\bibfnamefont {A.}~\bibnamefont
  {M\'{a}ttar}}, \bibinfo {author} {\bibfnamefont {P.}~\bibnamefont
  {Skrzypczyk}}, \bibinfo {author} {\bibfnamefont {E.}~\bibnamefont
  {Woodhead}}, \bibinfo {author} {\bibfnamefont {D.}~\bibnamefont
  {Cavalcanti}}, \bibinfo {author} {\bibfnamefont {K.}~\bibnamefont
  {Banaszek}},  and \bibinfo {author} {\bibfnamefont {A.}~\bibnamefont
  {Ac\'{\i}n}}, }\href {\doibase10.22331/q-2020-04-30-260} {\bibfield
  {journal} {\bibinfo  {journal} {\quantum} }\textbf {\bibinfo {volume} {4}},
  \bibinfo {pages} {260} (\bibinfo {year} {2020})}, \Eprint
  {http://arxiv.org/abs/1803.07089} {ar{X}iv:1803.07089 [quant-ph]}\BibitemShut
  {NoStop}%
\bibitem [{\citenamefont {Nieto~Silleras} \emph {et~al.}(2014)\citenamefont
  {Nieto~Silleras}, \citenamefont {Pironio}, and \citenamefont
  {Silman}}]{ref:nsps2014}%
  \BibitemOpen
  \bibfield  {author} {\bibinfo {author} {\bibfnamefont {O.}~\bibnamefont
  {Nieto~Silleras}}, \bibinfo {author} {\bibfnamefont {S.}~\bibnamefont
  {Pironio}},  and \bibinfo {author} {\bibfnamefont {J.}~\bibnamefont
  {Silman}}, }\href {\doibase10.1088/1367-2630/16/1/013035} {\bibfield
  {journal} {\bibinfo  {journal} {\njp} }\textbf {\bibinfo {volume} {16}},
  \bibinfo {pages} {013035} (\bibinfo {year} {2014})}, \Eprint
  {http://arxiv.org/abs/1309.3930} {ar{X}iv:1309.3930 [quant-ph]}\BibitemShut
  {NoStop}%
\bibitem [{\citenamefont {Bancal} \emph {et~al.}(2014)\citenamefont {Bancal},
  \citenamefont {Sheridan}, and \citenamefont {Scarani}}]{ref:bss2014}%
  \BibitemOpen
  \bibfield  {author} {\bibinfo {author} {\bibfnamefont {J.-D.} \bibnamefont
  {Bancal}}, \bibinfo {author} {\bibfnamefont {L.}~\bibnamefont {Sheridan}},
  and \bibinfo {author} {\bibfnamefont {V.}~\bibnamefont {Scarani}}, }\href
  {\doibase10.1088/1367-2630/16/3/033011} {\bibfield  {journal} {\bibinfo
  {journal} {\njp} }\textbf {\bibinfo {volume} {16}}, \bibinfo {pages} {033011}
  (\bibinfo {year} {2014})}, \Eprint {http://arxiv.org/abs/1309.3894}
  {ar{X}iv:1309.3894 [quant-ph]}\BibitemShut {NoStop}%
\bibitem [{\citenamefont {Lawson} \emph {et~al.}(2010)\citenamefont {Lawson},
  \citenamefont {Linden}, and \citenamefont {Popescu}}]{ref:llp2010}%
  \BibitemOpen
  \bibfield  {author} {\bibinfo {author} {\bibfnamefont {T.}~\bibnamefont
  {Lawson}}, \bibinfo {author} {\bibfnamefont {N.}~\bibnamefont {Linden}},  and
  \bibinfo {author} {\bibfnamefont {S.}~\bibnamefont {Popescu}}, }\href@noop {}
  {  (\bibinfo {year} {2010})}, \Eprint {http://arxiv.org/abs/1011.6245}
  {ar{X}iv:1011.6245 [quant-ph]}\BibitemShut {NoStop}%
\bibitem [{\citenamefont {Ac\'\i{}n} \emph {et~al.}(2012)\citenamefont
  {Ac\'\i{}n}, \citenamefont {Massar}, and \citenamefont
  {Pironio}}]{ref:amp2012}%
  \BibitemOpen
  \bibfield  {author} {\bibinfo {author} {\bibfnamefont {A.}~\bibnamefont
  {Ac\'\i{}n}}, \bibinfo {author} {\bibfnamefont {S.}~\bibnamefont {Massar}},
  and \bibinfo {author} {\bibfnamefont {S.}~\bibnamefont {Pironio}}, }\href
  {\doibase10.1103/PhysRevLett.108.100402} {\bibfield  {journal} {\bibinfo
  {journal} {\prl} }\textbf {\bibinfo {volume} {108}}, \bibinfo {pages}
  {100402} (\bibinfo {year} {2012})}, \Eprint {http://arxiv.org/abs/1107.2754}
  {ar{X}iv:1107.2754 [quant-ph]}\BibitemShut {NoStop}%
\bibitem [{\citenamefont {Woodhead} and \citenamefont
  {Pironio}(2015)}]{ref:wp2015}%
  \BibitemOpen
  \bibfield  {author} {\bibinfo {author} {\bibfnamefont {E.}~\bibnamefont
  {Woodhead}} and \bibinfo {author} {\bibfnamefont {S.}~\bibnamefont
  {Pironio}}, }\href {\doibase10.1103/PhysRevLett.115.150501} {\bibfield
  {journal} {\bibinfo  {journal} {\prl} }\textbf {\bibinfo {volume} {115}},
  \bibinfo {pages} {150501} (\bibinfo {year} {2015})}, \Eprint
  {http://arxiv.org/abs/1507.02889} {ar{X}iv:1507.02889 [quant-ph]}\BibitemShut
  {NoStop}%
\bibitem [{\citenamefont {Woodhead}(2014{\natexlab{a}})}]{ref:w2014b}%
  \BibitemOpen
  \bibfield  {author} {\bibinfo {author} {\bibfnamefont {E.}~\bibnamefont
  {Woodhead}}, }\emph {\bibinfo {title} {Imperfections and self testing in
  prepare-and-measure quantum key distribution}}, \href
  {https://difusion.ulb.ac.be/vufind/Record/ULB-DIPOT:oai:dipot.ulb.ac.be:2013/209185/Holdings}
  {Ph.D. thesis}, \bibinfo  {school} {Universit\'e{} libre de Bruxelles}
  (\bibinfo {year} {2014}{\natexlab{a}})\BibitemShut {NoStop}%
\bibitem [{\citenamefont {Berta} \emph {et~al.}(2010)\citenamefont {Berta},
  \citenamefont {Christandl}, \citenamefont {Colbeck}, \citenamefont {Renes},
  and \citenamefont {Renner}}]{ref:bc2010}%
  \BibitemOpen
  \bibfield  {author} {\bibinfo {author} {\bibfnamefont {M.}~\bibnamefont
  {Berta}}, \bibinfo {author} {\bibfnamefont {M.}~\bibnamefont {Christandl}},
  \bibinfo {author} {\bibfnamefont {R.}~\bibnamefont {Colbeck}}, \bibinfo
  {author} {\bibfnamefont {J.~M.} \bibnamefont {Renes}},  and \bibinfo {author}
  {\bibfnamefont {R.}~\bibnamefont {Renner}}, }\href
  {\doibase10.1038/nphys1734} {\bibfield  {journal} {\bibinfo  {journal}
  {\natphys} }\textbf {\bibinfo {volume} {6}}, \bibinfo {pages} {659} (\bibinfo
  {year} {2010})}, \Eprint {http://arxiv.org/abs/0909.0950} {ar{X}iv:0909.0950
  [quant-ph]}\BibitemShut {NoStop}%
\bibitem [{\citenamefont {Woodhead}(2016)}]{ref:w2016}%
  \BibitemOpen
  \bibfield  {author} {\bibinfo {author} {\bibfnamefont {E.}~\bibnamefont
  {Woodhead}}, }\href {\doibase10.1088/1367-2630/18/5/055010} {\bibfield
  {journal} {\bibinfo  {journal} {\njp} }\textbf {\bibinfo {volume} {18}},
  \bibinfo {pages} {055010} (\bibinfo {year} {2016})}, \Eprint
  {http://arxiv.org/abs/1512.03387} {ar{X}iv:1512.03387 [quant-ph]}\BibitemShut
  {NoStop}%
\bibitem [{\citenamefont {Woodhead}(2013)}]{ref:w2013}%
  \BibitemOpen
  \bibfield  {author} {\bibinfo {author} {\bibfnamefont {E.}~\bibnamefont
  {Woodhead}}, }\href {\doibase10.1103/PhysRevA.88.012331} {\bibfield
  {journal} {\bibinfo  {journal} {\pra} }\textbf {\bibinfo {volume} {88}},
  \bibinfo {pages} {012331} (\bibinfo {year} {2013})}, \Eprint
  {http://arxiv.org/abs/1303.4821} {ar{X}iv:1303.4821 [quant-ph]}\BibitemShut
  {NoStop}%
\bibitem [{\citenamefont {Jordan}(1875)}]{ref:j1875}%
  \BibitemOpen
  \bibfield  {author} {\bibinfo {author} {\bibfnamefont {C.}~\bibnamefont
  {Jordan}}, }\href {http://www.numdam.org/item?id=BSMF_1875__3__103_2}
  {\bibfield  {journal} {\bibinfo  {journal} {\bsmf} }\textbf {\bibinfo
  {volume} {3}}, \bibinfo {pages} {103} (\bibinfo {year} {1875})}\BibitemShut
  {NoStop}%
\bibitem [{\citenamefont {Tsirelson}(1993)}]{ref:t1993}%
  \BibitemOpen
  \bibfield  {author} {\bibinfo {author} {\bibfnamefont {B.~S.} \bibnamefont
  {Tsirelson}}, }\href {https://www.tau.ac.il/~tsirel/download/hadron.html}
  {\bibfield  {journal} {\bibinfo  {journal} {\hjsupp} }\textbf {\bibinfo
  {volume} {8}}, \bibinfo {pages} {329} (\bibinfo {year} {1993})}\BibitemShut
  {NoStop}%
\bibitem [{\citenamefont {Masanes}(2006)}]{ref:m2006}%
  \BibitemOpen
  \bibfield  {author} {\bibinfo {author} {\bibfnamefont {L.}~\bibnamefont
  {Masanes}}, }\href {\doibase10.1103/PhysRevLett.97.050503} {\bibfield
  {journal} {\bibinfo  {journal} {\prl} }\textbf {\bibinfo {volume} {97}},
  \bibinfo {pages} {050503} (\bibinfo {year} {2006})}, \Eprint
  {http://arxiv.org/abs/quant-ph/0512153}
  {ar{X}iv:quant-ph/0512153}\BibitemShut {NoStop}%
\bibitem [{\citenamefont {Woodhead}(2014{\natexlab{b}})}]{ref:w2014}%
  \BibitemOpen
  \bibfield  {author} {\bibinfo {author} {\bibfnamefont {E.}~\bibnamefont
  {Woodhead}}, }\href {\doibase10.1103/PhysRevA.90.022306} {\bibfield
  {journal} {\bibinfo  {journal} {\pra} }\textbf {\bibinfo {volume} {90}},
  \bibinfo {pages} {022306} (\bibinfo {year} {2014}{\natexlab{b}})}, \Eprint
  {http://arxiv.org/abs/1405.5625} {ar{X}iv:1405.5625 [quant-ph]}\BibitemShut
  {NoStop}%
\bibitem [{\citenamefont {Fuchs} \emph {et~al.}(1997)\citenamefont {Fuchs},
  \citenamefont {Gisin}, \citenamefont {Griffiths}, \citenamefont {Niu}, and
  \citenamefont {Peres}}]{ref:fg1997}%
  \BibitemOpen
  \bibfield  {author} {\bibinfo {author} {\bibfnamefont {C.~A.} \bibnamefont
  {Fuchs}}, \bibinfo {author} {\bibfnamefont {N.}~\bibnamefont {Gisin}},
  \bibinfo {author} {\bibfnamefont {R.~B.} \bibnamefont {Griffiths}}, \bibinfo
  {author} {\bibfnamefont {C.-S.} \bibnamefont {Niu}},  and \bibinfo {author}
  {\bibfnamefont {A.}~\bibnamefont {Peres}}, }\href
  {\doibase10.1103/PhysRevA.56.1163} {\bibfield  {journal} {\bibinfo  {journal}
  {\pra} }\textbf {\bibinfo {volume} {56}}, \bibinfo {pages} {1163} (\bibinfo
  {year} {1997})}, \Eprint {http://arxiv.org/abs/quant-ph/9701039}
  {ar{X}iv:quant-ph/9701039}\BibitemShut {NoStop}%
\bibitem [{\citenamefont {Eberhard}(1993)}]{ref:e1993}%
  \BibitemOpen
  \bibfield  {author} {\bibinfo {author} {\bibfnamefont {P.~H.} \bibnamefont
  {Eberhard}}, }\href {\doibase10.1103/PhysRevA.47.R747} {\bibfield  {journal}
  {\bibinfo  {journal} {\pra} }\textbf {\bibinfo {volume} {47}}, \bibinfo
  {pages} {R747} (\bibinfo {year} {1993})}\BibitemShut {NoStop}%
\bibitem [{\citenamefont {Ma} and \citenamefont
  {Lutkenhaus}(2012)}]{ref:ml2012}%
  \BibitemOpen
  \bibfield  {author} {\bibinfo {author} {\bibfnamefont {X.}~\bibnamefont {Ma}}
  and \bibinfo {author} {\bibfnamefont {N.}~\bibnamefont {Lutkenhaus}}, }\href
  {\doibase10.26421/QIC12.3-4-2} {\bibfield  {journal} {\bibinfo  {journal}
  {\quantic} }\textbf {\bibinfo {volume} {12}}, \bibinfo {pages} {203}
  (\bibinfo {year} {2012})}, \Eprint {http://arxiv.org/abs/1109.1203}
  {ar{X}iv:1109.1203 [quant-ph]}\BibitemShut {NoStop}%
\bibitem [{\citenamefont {Lasserre}(2001)}]{ref:l2001}%
  \BibitemOpen
  \bibfield  {author} {\bibinfo {author} {\bibfnamefont {J.~B.} \bibnamefont
  {Lasserre}}, }\href {\doibase10.1137/S1052623400366802} {\bibfield  {journal}
  {\bibinfo  {journal} {\siamjc} }\textbf {\bibinfo {volume} {11}}, \bibinfo
  {pages} {796} (\bibinfo {year} {2001})}\BibitemShut {NoStop}%
\bibitem [{\citenamefont {Ho} \emph {et~al.}(2020)\citenamefont {Ho},
  \citenamefont {Sekatski}, \citenamefont {Tan}, \citenamefont {Renner},
  \citenamefont {Bancal}, and \citenamefont {Sangouard}}]{ref:hs2020}%
  \BibitemOpen
  \bibfield  {author} {\bibinfo {author} {\bibfnamefont {M.}~\bibnamefont
  {Ho}}, \bibinfo {author} {\bibfnamefont {P.}~\bibnamefont {Sekatski}},
  \bibinfo {author} {\bibfnamefont {E.~Y.-Z.} \bibnamefont {Tan}}, \bibinfo
  {author} {\bibfnamefont {R.}~\bibnamefont {Renner}}, \bibinfo {author}
  {\bibfnamefont {J.-D.} \bibnamefont {Bancal}},  and \bibinfo {author}
  {\bibfnamefont {N.}~\bibnamefont {Sangouard}}, }\href
  {\doibase10.1103/PhysRevLett.124.230502} {\bibfield  {journal} {\bibinfo
  {journal} {\prl} }\textbf {\bibinfo {volume} {124}}, \bibinfo {pages}
  {230502} (\bibinfo {year} {2020})}, \Eprint {http://arxiv.org/abs/2005.13015}
  {ar{X}iv:2005.13015 [quant-ph]}\BibitemShut {NoStop}%
\bibitem [{\citenamefont {Sekatski} \emph {et~al.}(2020)\citenamefont
  {Sekatski}, \citenamefont {Bancal}, \citenamefont {Valcarce}, \citenamefont
  {Tan}, \citenamefont {Renner}, and \citenamefont {Sangouard}}]{ref:sb2020}%
  \BibitemOpen
  \bibfield  {author} {\bibinfo {author} {\bibfnamefont {P.}~\bibnamefont
  {Sekatski}}, \bibinfo {author} {\bibfnamefont {J.-D.} \bibnamefont {Bancal}},
  \bibinfo {author} {\bibfnamefont {X.}~\bibnamefont {Valcarce}}, \bibinfo
  {author} {\bibfnamefont {E.~Y.-Z.} \bibnamefont {Tan}}, \bibinfo {author}
  {\bibfnamefont {R.}~\bibnamefont {Renner}},  and \bibinfo {author}
  {\bibfnamefont {N.}~\bibnamefont {Sangouard}}, }\href@noop {} {  (\bibinfo
  {year} {2020})}, \Eprint {http://arxiv.org/abs/2009.01784}
  {ar{X}iv:2009.01784 [quant-ph]}\BibitemShut {NoStop}%
\bibitem [{\citenamefont {Bhavsar} \emph {et~al.}(2021)\citenamefont {Bhavsar},
  \citenamefont {Ragy}, and \citenamefont {Colbeck}}]{ref:brc2021}%
  \BibitemOpen
  \bibfield  {author} {\bibinfo {author} {\bibfnamefont {R.}~\bibnamefont
  {Bhavsar}}, \bibinfo {author} {\bibfnamefont {S.}~\bibnamefont {Ragy}},  and
  \bibinfo {author} {\bibfnamefont {R.}~\bibnamefont {Colbeck}}, }\href@noop {}
  {  (\bibinfo {year} {2021})}, \Eprint {http://arxiv.org/abs/2103.07504}
  {ar{X}iv:2103.07504 [quant-ph]}\BibitemShut {NoStop}%
\end{thebibliography}%

\appendix

\clearpage

\section{Qubit optimisation problem}
\label{sec:solution}

Here we solve the optimisation problem \eqref{eq:avgXB-optimisation} in
section~\ref{sec:derivation} of the main text. We first simplify it by
introducing polar coordinates,
\begin{IEEEeqnarray}{rCl+rCl}
  E_{\rzz} &=& \lambda \cos(z) \,, &
  E_{\rxz} &=& \mu \cos(x) \,, \\
  E_{\rzx} &=& \lambda \sin(z) \,, &
  E_{\rxx} &=& \mu \sin(x) \,.
\end{IEEEeqnarray}
With this change of variables the problem becomes
\begin{IEEEeqnarray}{u+l}
  \label{eq:problem-polar-coords}
  minimise & \abs{\mu} \nonumber \\
  subject to & \left\{
    \begin{IEEEeqnarraybox}[][c]{rCl}
      \alpha^{2} \lambda^{2} \cos(z)^{2} + \lambda^{2} \sin(z)^{2} && \\
      +\> \mu^{2} \sin(x)^{2}
      &\geq& S\du{\alpha}{2}/4 \\
      \lambda^{2} &\leq& 1 \\
      \mu^{2} &\leq& 1 \\
      (1 - \lambda^{2}) (1 - \mu^{2}) && \\
      -\> \lambda^{2} \mu^{2} \cos(z - x)^{2}
      &\geq& 0
    \end{IEEEeqnarraybox} \right. \IEEEeqnarraynumspace
\end{IEEEeqnarray}
in the free variables $\mu$, $\lambda$, $z$, and $x$. From here, it is an
algebra problem to eliminate the unwanted variables $\lambda$, $z$, and $x$
so that only a constraint between $\abs{\mu}$ and the constants $\alpha$ and
$S_{\alpha}$ remains.

We begin with the first constraint. Using the trigonometric identities
$\cos(x)^{2} = \bigro{1 + \cos(2x)}/2$ and
$\sin(x)^{2} = \bigro{1 - \cos(2x)}/2$, it can be rewritten
\begin{IEEEeqnarray}{rCl}
  (1 + \alpha^{2}) \lambda^{2} + \mu^{2}
  - (1 - \alpha^{2}) \lambda^{2} \cos(2z) && \nonumber \\
  -\> \mu^{2} \cos(2x)
  &\geq& S\du{\alpha}{2} / 2 \,. \IEEEeqnarraynumspace
\end{IEEEeqnarray}
Substituting $\Sigma = z + x$ and $\Delta = z - x$ and using the
trigonometric angle sum and difference identities, we get
\begin{IEEEeqnarray}{rCl}
  (1 + \alpha^{2}) \lambda^{2} + \mu^{2} && \nonumber \\
  - \bigro{(1 - \alpha^{2}) \lambda^{2} + \mu^{2}}
  \cos(\Sigma) \cos(\Delta) && \nonumber \\
  -\> \bigro{(1 - \alpha^{2}) \lambda^{2} - \mu^{2}}
  \sin(\Sigma) \sin(\Delta)
  &\geq& S\du{\alpha}{2} / 2 \,. \IEEEeqnarraynumspace
\end{IEEEeqnarray}
We can then maximise the left-hand side over $\Sigma$, which doesn't appear
in any of the other constraints. After simplifying a little this gives
\begin{IEEEeqnarray}{l}
  \label{eq:lambda-mu-sqrt-delta-constraint}
  (1 + \alpha^{2}) \lambda^{2} + \mu^{2} \nonumber \\
  +\> \sqrt{\bigro{(1 - \alpha^{2}) \lambda^{2} - \mu^{2}}^{2}
    + 4 (1 - \alpha^{2}) \lambda^{2} \mu^{2} \cos(\Delta)^{2}}
  \nonumber \\
  \qquad \geq S\du{\alpha}{2} / 2 \,,
\end{IEEEeqnarray}
which we can rearrange to
\begin{IEEEeqnarray}{l}
  \label{eq:sqrt-lambda-mu-delta-constraint}
  \sqrt{\bigro{(1 - \alpha^{2}) \lambda^{2} - \mu^{2}}^{2}
    + 4 (1 - \alpha^{2}) \lambda^{2} \mu^{2} \cos(\Delta)^{2}}
  \nonumber \\
  \qquad \geq \bigro{(1 - \alpha^{2}) \lambda^{2} - \mu^{2}}
  + 2 (S\du{\alpha}{2} / 4 - \lambda^{2}) \,.
\end{IEEEeqnarray}
Now note that, if $\alpha^{2} < 1$, the term
$(1 - \alpha^{2}) \lambda^{2} \mu^{2} \cos(\Delta)^{2}$ is nonnegative. Hence
we also have
\begin{IEEEeqnarray}{rCl}
  \IEEEeqnarraymulticol{3}{l}{
    \label{eq:sqrt-lambda-mu-delta-negative}
    \sqrt{\bigro{(1 - \alpha^{2}) \lambda^{2} - \mu^{2}}^{2}
      + 4 (1 - \alpha^{2}) \lambda^{2} \mu^{2} \cos(\Delta)^{2}}}
  \nonumber \\
  \qquad &\geq& \babs{(1 - \alpha^{2}) \lambda^{2} - \mu^{2}}
  \nonumber \\
  &\geq& - \bigro{(1 - \alpha^{2}) \lambda^{2} - \mu^{2}} \nonumber \\
  &\geq& - \bigro{(1 - \alpha^{2}) \lambda^{2} - \mu^{2}}
  - 2 (S\du{\alpha}{2}/4 - \lambda^{2}) \,, \IEEEeqnarraynumspace
\end{IEEEeqnarray}
where we assume that $\abs{S_{\alpha}} \geq 2$ (i.e., $S_{\alpha}$ is
attaining or exceeding the classical bound) and $\lambda^{2} \leq 1$ (which
is one of the problem constraints in \eqref{eq:problem-polar-coords}), which
together imply $S\du{\alpha}{2}/4 - \lambda^{2} \geq 0$, in order to get the
last line. Eqs. \eqref{eq:sqrt-lambda-mu-delta-constraint} and
\eqref{eq:sqrt-lambda-mu-delta-negative} together confirm that
\begin{IEEEeqnarray}{l}
  \label{eq:sqrt-lambda-mu-delta-abs}
  \sqrt{\bigro{(1 - \alpha^{2}) \lambda^{2} - \mu^{2}}^{2}
    + 4 (1 - \alpha^{2}) \lambda^{2} \mu^{2} \cos(\Delta)^{2}}
  \IEEEeqnarraynumspace \nonumber \\
  \qquad \geq \babs{\bigro{(1 - \alpha^{2}) \lambda^{2} - \mu^{2}}
    + 2 (S\du{\alpha}{2}/4 - \lambda^{2})}
\end{IEEEeqnarray}
holds with the absolute value term on the right. Now, since the right side of
\eqref{eq:sqrt-lambda-mu-delta-abs} is nonnegative we are justified to square
both sides of the inequality. After doing this and simplifying the result, we
obtain
\begin{IEEEeqnarray}{l}
  \label{eq:lambda-mu-delta-after-squaring}
  (1 - \alpha^{2}) \lambda^{2} \mu^{2} \cos(\Delta)^{2} \nonumber \\
  \qquad \geq (S\du{\alpha}{2}/4 - \alpha^{2} \lambda^{2} - \mu^{2})
  (S\du{\alpha}{2}/4 - \lambda^{2}) \,. \IEEEeqnarraynumspace
\end{IEEEeqnarray}

We now eliminate $\Delta$ from the problem by applying the constraint
\begin{equation}
  (1 - \lambda^{2})(1 - \mu^{2}) \geq \lambda^{2} \mu^{2} \cos(\Delta)^{2}
\end{equation}
to the left side of \eqref{eq:lambda-mu-delta-after-squaring}, obtaining
\begin{IEEEeqnarray}{l}
  \label{eq:lambda-mu-constraint}
  (1 - \alpha^{2}) (1 - \lambda^{2}) (1 - \mu^{2}) \nonumber \\
  \qquad \geq (S\du{\alpha}{2}/4 - \alpha^{2} \lambda^{2} - \mu^{2})
  (S\du{\alpha}{2}/4 - \lambda^{2}) \,. \IEEEeqnarraynumspace
\end{IEEEeqnarray}
Collecting the terms in $\mu^{2}$ together we can rewrite
\eqref{eq:lambda-mu-constraint} as
\begin{equation}
  \label{eq:collected-mu2-terms}
  (X + \alpha^{2} \Lambda) \mu^{2}
  \geq X + (X + \alpha^{2} \Lambda) \Lambda
\end{equation}
where
\begin{IEEEeqnarray}{rCl}
  X &=& (1 - \alpha^{2}) (S\du{\alpha}{2}/4 - 1) \,, \\
  \Lambda &=& S\du{\alpha}{2}/4 - \lambda^{2} \,.
\end{IEEEeqnarray}
Note that here both $X$ and $\Lambda$ are strictly positive assuming
$\lambda^{2} \leq 1$, $\abs{S}_{\alpha} > 2$, and $\abs{\alpha} < 1$, and
only $\Lambda$ depends on the remaining parameter $\lambda$. Subject to these
conditions, \eqref{eq:collected-mu2-terms} gives a lower bound for $\mu^{2}$
in terms of $\lambda$ which we can express as
\begin{equation}
  \label{eq:mu2-lambda-constraint}
  \mu^{2} \geq \frac{\sqrt{X}}{\abs{\alpha}}
  \, f \biggro{\frac{X + \alpha^{2} \Lambda}{\abs{\alpha} \sqrt{X}}}
  - \frac{X}{\alpha^{2}} \,,
\end{equation}
where we have made appear the function
\begin{equation}
  \label{eq:t-def}
  f(t) = t + 1/t \,.
\end{equation}

The remaining problem is to minimise the right side of
\eqref{eq:mu2-lambda-constraint} subject to the condition
$\lambda^{2} \leq 1$. This is straightforward due to the characteristics of
the function $f$ \eqref{eq:t-def} that we expressed it in terms of: for
$t > 0$, $t$ is convex and its global minimum of $f(t) = 2$ is attained at
$t = 1$, so the lower bound for $\mu^{2}$ is determined by how close we can
make the argument
\begin{equation}
  t = \frac{X + \alpha^{2} \Lambda}{\abs{\alpha} \sqrt{X}}
\end{equation}
to $1$. The limits $0 \leq \lambda^{2} \leq 1$ translate to
\begin{equation}
  \label{eq:arg-upper}
  t \leq \frac{\alpha^{2} + S\du{\alpha}{2}/4 - 1}{
    \sqrt{\alpha^{2} (1 - \alpha^{2})(S\du{\alpha}{2}/4 - 1)}}
\end{equation}
and
\begin{equation}
  \label{eq:arg-lower}
  t
  \geq \sqrt{\frac{S\du{\alpha}{2}/4 - 1}{\alpha^{2} (1 - \alpha^{2})}} \,.
\end{equation}
The upper limit \eqref{eq:arg-upper} can be rewritten as
\begin{equation}
  t \leq \sqrt{1
    + \frac{\alpha^{2} \bigro{(1 + \alpha^{2}) S\du{\alpha}{2}/4 - 1}
      + (S\du{\alpha}{2}/4 - 1)^{2}}
    {\alpha^{2} (1 - \alpha^{2}) (S\du{\alpha}{2}/4 - 1)}} \,,
\end{equation}
which makes it clear that the right side is never less than $1$. The
lower limit \eqref{eq:arg-lower} on the other hand may be less than $1$
depending on $\alpha$ and $S_{\alpha}$. Specifically, if
\begin{equation}
  \abs{S_{\alpha}} \leq 2 \sqrt{1 + \alpha^{2} - \alpha^{4}}
\end{equation}
then the right side of \eqref{eq:arg-lower} is not more than $1$, in which
case it is possible to choose $\lambda$ such that $t = 1$. Recalling that
$\abs{\mu} = \abs{\avg{\sx \otimes B}}$, we obtain in this case
\begin{equation}
  \avg{\sx \otimes B}^{2}
  \geq 1 - \Bigro{1 - \tfrac{1}{\abs{\alpha}}
    \sqrt{(1 - \alpha^{2}) (S\du{\alpha}{2}/4 - 1)}}^{2} \,.
\end{equation}
On the other hand, if
$\abs{S_{\alpha}} \geq 2 \sqrt{1 + \alpha^{2} - \alpha^{4}}$ then the minimum
is attained with the smallest value allowed of the argument,
\begin{equation}
  t = \sqrt{\frac{S\du{\alpha}{2}/4 - 1}{\alpha^{2} (1 - \alpha^{2})}}
  = \frac{\sqrt{X}}{\abs{\alpha} (1 - \alpha^{2})} \,,
\end{equation}
in which case the constraint \eqref{eq:mu2-lambda-constraint} simplifies to
the same expression,
\begin{equation}
  \label{eq:corr-bound}
  \abs{\avg{\sx \otimes B}} \geq \sqrt{S\du{\alpha}{2}/4 - \alpha^{2}} \,,
\end{equation}
that we derived in the main text for $\abs{\alpha} \geq 1$.

\section{Concav/exity of the qubit bound}
\label{sec:concavexity}

Here we prove lemmas~\ref{lem:convexity} and \ref{lem:concavity} in the main
text by bounding the first and second derivatives of the functions they
concern. Both of these depend on the function
\begin{equation}
  \phi(x)
  = 1 - \tfrac{1}{2} (1 + x) \log(1 + x)
  - \tfrac{1}{2} (1 - x) \log(1 - x)
\end{equation}
that we defined in the introduction. We give its first and
second derivatives, which are used in the proofs, here for convenience:
\begin{IEEEeqnarray}{rCl}
  \phi'(x) &=& - \frac{1}{2} \log_{2} \biggro{\frac{1 + x}{1 - x}} \,, \\
  \phi''(x) &=& - \frac{1}{\log(2)} \frac{1}{1 - x^{2}} \,.
\end{IEEEeqnarray}

\subsection{Proof of lemma~\ref{lem:convexity}}
\label{sec:convexity}

We express the function $f$ defined in Eq.~\eqref{eq:f-convexity} as
\begin{equation}
  f(x) = 1 + \phi(R) - \phi(r)
\end{equation}
with $R = \sqrt{Q + (1 - Q) x}$ and $r = \sqrt{x}$. The first and second
derivatives of $r$ and $R$ are
\begin{IEEEeqnarray}{rCl+rCl}
  r' &=& \frac{1}{2 r} \,, & r'' &=& - \frac{1}{4 r^{3}} \,, \\
  R' &=& \frac{1 - Q}{2 R} \,, &
  R'' &=& - \frac{(1 - Q)^{2}}{4 R^{3}} \,. \IEEEeqnarraynumspace
\end{IEEEeqnarray}

Let us first verify that $f$ is monotonically increasing. Its first
derivative is
\begin{IEEEeqnarray}{rClll}
  f'(x) &=& \IEEEeqnarraymulticol{3}{l}{
    \phi'(R) R' - \phi'(r) r'} \nonumber \\
  &=& \frac{1}{2 \log(2)} \Biggsq{&
    - \frac{1 - Q}{2 R} \log \biggro{\frac{1 + R}{1 - R}} & \nonumber \\
    &&&+\> \frac{1}{2 r} \log \biggro{\frac{1 + r}{1 - r}} &} \,.
\end{IEEEeqnarray}
To change the terms with $\log$s into something easier to work with we
substitute
\begin{equation}
  \label{eq:integral-trick}
  \frac{1}{2 \xi} \log \biggro{\frac{1 + \xi}{1 - \xi}}
  = \int_{0}^{1} \dd u \, \frac{1}{1 - \xi^{2} u^{2}}
\end{equation}
for both $\xi = R$ and $\xi = r$. The rest of the proof then amounts to
manipulating and simplifying quotients of polynomials:
\begin{IEEEeqnarray}{rCl}
  f'(x) &=& \frac{1}{2 \log(2)} \int_{0}^{1} \dd u \, \biggsq{
    - \frac{1 - Q}{1 - R^{2} u^{2}}
    + \frac{1}{1 - r^{2} u^{2}}} \nonumber \\
  &=& \frac{1}{2 \log(2)} \int_{0}^{1} \dd u \,
  \frac{Q (1 - r^{2} u^{2}) - (R^{2} - r^{2}) u^{2}}
  {(1 - r^{2} u^{2})(1 - R^{2} u^{2})} \nonumber \\
  &=& \frac{Q}{2 \log(2)} \int_{0}^{1} \dd u \,
  \frac{1 - u^{2}}{(1 - r^{2} u^{2})(1 - R^{2} u^{2})} \nonumber \\
  &\geq& 0 \,,
\end{IEEEeqnarray}
where we used that $R^{2} - r^{2} = Q (1 - r^{2})$.

We prove that $f$ is convex in a similar way. Its second derivative is
\begin{IEEEeqnarray}{rCl}
  f''(x) &=& \phi''(R) R'^{2} - \phi''(r) r'^{2} \nonumber \\
  &&+\> \phi'(R) R'' - \phi'(r) r'' \,.
\end{IEEEeqnarray}
The first and second lines on the right side evaluate to
\begin{IEEEeqnarray}{rCl}
  \IEEEeqnarraymulticol{3}{l}{
    \label{eq:fphi2-diff}
    \phi''(R) R'^{2} - \phi''(r) r'^{2}} \nonumber \\
  \quad &=& \frac{1}{4 \log(2)} \biggsq{
    - \frac{(1 - Q)^{2}}{R^{2} (1 - R^{2})}
    + \frac{1}{r^{2} (1 - r^{2})}} \IEEEeqnarraynumspace
\end{IEEEeqnarray}
and
\begin{IEEEeqnarray}{rCl}
  \IEEEeqnarraymulticol{3}{l}{\label{eq:fphi1-diff}
    \phi'(R) R'' - \phi'(r) r''} \nonumber \\
  \; &=&
  \frac{1}{4 \log(2)}
  \begin{IEEEeqnarraybox}[][t]{rll}
    \Biggsq{&
      \frac{(1 - Q)^{2}}{R^{2}} \, \frac{1}{2 R}
      \log \biggro{\frac{1 + R}{1 - R}} & \\
      &-\> \frac{1}{r^{2}} \, \frac{1}{2 r}
      \log \biggro{\frac{1 + r}{1 - r}} &}
  \end{IEEEeqnarraybox} \nonumber \\
  &=& \frac{1}{4 \log(2)} \int_{0}^{1} \dd u
  \biggsq{
    \frac{(1 - Q)^{2}}{R^{2} (1 - R^{2} u^{2})}
    - \frac{1}{r^{2} (1 - r^{2} u^{2})}} \,. \nonumber \\*
\end{IEEEeqnarray}
Adding \eqref{eq:fphi2-diff} and \eqref{eq:fphi1-diff} and using that
\begin{equation}
  \frac{1}{\xi^{2} (1 - \xi^{2})} - \frac{1}{\xi^{2} (1 - \xi^{2} u^{2})}
  = \frac{(1 - u^{2})}{(1 - \xi^{2}) (1 - \xi^{2} u^{2})}
\end{equation}
for $\xi = r$ and $\xi = R$ and that
\begin{equation}
  1 - R^{2} = (1 - Q) (1 - r^{2})
\end{equation}
we get
\begin{IEEEeqnarray}{rCl}
  f''(x) &=& \frac{1}{4 \log(2)} \int_{0}^{1} \dd u
  \begin{IEEEeqnarraybox}[][t]{rll}
    \Biggsq{&
      - \frac{(1 - Q)^{2} (1 - u^{2})}{(1 - R^{2}) (1 - R^{2} u^{2})} & \\
      &+\> \frac{1 - u^{2}}{(1 - r^{2}) (1 - r^{2} u^{2})} &}
  \end{IEEEeqnarraybox} \nonumber \\
  &=& \frac{1}{4 \log(2)} \int_{0}^{1} \dd u \,
  \begin{IEEEeqnarraybox}[][t]{l}
    \frac{1 - u^{2}}{1 - r^{2}} \\
    \times\> \biggsq{- \frac{1 - Q}{1 - R^{2} u^{2}}
      + \frac{1}{1 - r^{2} u^{2}}}
  \end{IEEEeqnarraybox}
  \nonumber \\
  &=& \frac{1}{4 \log(2)} \, \frac{Q}{1 - r^{2}}
  \int_{0}^{1} \dd u \, \frac{(1 - u^{2})^{2}}{
    (1 - r^{2} u^{2}) (1 - R^{2} u^{2})} \nonumber \\
  &\geq& 0 \,.
\end{IEEEeqnarray}

\subsection{Proof of lemma~\ref{lem:concavity}}
\label{sec:concavity}

We proceed similarly to the proof of lemma~\ref{lem:convexity}. We write the
function $f$ defined in \eqref{eq:f-concavity} as
\begin{equation}
  f(x) = 1 + \phi(R) - \phi(r) \,,
\end{equation}
this time with $r = \sqrt{1 - x^{2}}$ and $R = \sqrt{Q + (1 - Q) r^{2}}$, for
which
\begin{IEEEeqnarray}{rCl+rCl}
  r' &=& -\frac{\sqrt{1 - r^{2}}}{r} \,, &
  r'' &=& -\frac{1}{r^{3}} \,, \\
  R' &=& -\frac{\sqrt{(1 - Q)(1 - R^{2})}}{R} \,, &
  R'' &=& -\frac{1 - Q}{R^{3}} \,. \IEEEeqnarraynumspace
\end{IEEEeqnarray}

The first derivative of $f$ is
\begin{IEEEeqnarray}{rCl}
  f'(x) &=& \phi'(R) R' - \phi'(r) r' \nonumber \\
  &=& \frac{1}{\log(2)}
  \begin{IEEEeqnarraybox}[][t]{rll}
    \Biggsq{&
      \frac{\sqrt{(1 - Q) (1 - R^{2})}}{2 R}
      \log \biggro{\frac{1 + R}{1 - R}} & \\
      &-\> \frac{\sqrt{1 - r^{2}}}{2 r}
      \log \biggro{\frac{1 + r}{1 - r}} &}
  \end{IEEEeqnarraybox} \nonumber \\
  &=& \frac{1}{\log(2)} \int_{0}^{1} \dd u
  \begin{IEEEeqnarraybox}[][t]{rll}
    \Biggsq{&
      \frac{\sqrt{(1 - Q) (1 - R^{2})}}{1 - R^{2} u^{2}} & \\
      &-\> \frac{\sqrt{1 - r^{2}}}{1 - r^{2} u^{2}} &}
  \end{IEEEeqnarraybox} \nonumber \\
  &=& \frac{\sqrt{1 - r^{2}}}{\log(2)}
  \int_{0}^{1} \dd u \biggsq{
    \frac{1 - Q}{1 - R^{2} u^{2}} - \frac{1}{1 - r^{2} u^{2}}} \nonumber \\
  &=& - \frac{Q \sqrt{1 - r^{2}}}{\log(2)}
  \int_{0}^{1} \dd u \,
  \frac{1 - u^{2}}{(1 - r^{2} u^{2}) (1 - R^{2} u^{2})} \nonumber \\
  &\leq& 0 \,,
\end{IEEEeqnarray}
where we used that $\sqrt{1 - R^{2}} = \sqrt{(1 - Q) (1 - r^{2})}$ to get to
the fourth line.

The second derivative of $f$ is
\begin{IEEEeqnarray}{rCl}
  f''(x) &=& \phi''(R) R'^{2} + \phi'(R) R''
  - \phi''(r) r'^{2} - \phi'(r) r'' \nonumber \\
  &=& \frac{1}{\log(2)}
  \begin{IEEEeqnarraybox}[][t]{rll}
    \Biggsq{&- \frac{1 - Q}{R^{2}}
      + \frac{1 - Q}{2 R^{3}} \log \biggro{\frac{1 + R}{1 - R}} & \\
      &+\> \frac{1}{r^{2}}
      - \frac{1}{2 r^{3}} \log \biggro{\frac{1 + r}{1 - r}} &}
  \end{IEEEeqnarraybox} \nonumber \\
  &=& \frac{1}{\log(2)} \int_{0}^{1} \dd u
  \begin{IEEEeqnarraybox}[][t]{rll}
    \Biggsq{&
      -\frac{1 - Q}{R^{2}} \biggro{1 - \frac{1}{1 - R^{2} u^{2}}} & \\
      &+\> \frac{1}{r^{2}} \biggro{1 - \frac{1}{1 - r^{2} u^{2}}} &}
  \end{IEEEeqnarraybox} \nonumber \\
  &=& \frac{1}{\log(2)} \int_{0}^{1} \dd u
  \biggsq{
    \frac{(1 - Q) u^{2}}{1 - R^{2} u^{2}}
    - \frac{u^{2}}{1 - r^{2} u^{2}}} \nonumber \\
  &=& -\frac{Q}{\log(2)} \int_{0}^{1} \dd u \,
  \frac{u^{2} (1 - u^{2})}{(1 - r^{2} u^{2}) (1 - R^{2} u^{2})}
  \IEEEeqnarraynumspace \nonumber \\
  &\leq& 0 \,.
\end{IEEEeqnarray}

\section{Maximal noise preprocessing}
\label{sec:max-preprocessing}

Thresholds to the Devetak-Winter rate can be computed accurately in the limit
$q \to 1/2$ of maximal noise preprocessing by setting
$q = (1 - \varepsilon)/2$ and then expanding the expression for the key rate
to the first nontrivial power in $\varepsilon$ \cite{ref:w2014}. For the BB84
bound \eqref{eq:HZE-bb84-noise} the result is
\begin{equation}
  H(\sz | \rE)
  \gtrapprox 1 - \frac{1 - \avg{\sx \otimes B}^{2}}
  {4 \abs{\avg{\sx \otimes B}}}
  \log_{2} \biggro{
    \frac{1 + \abs{\avg{\sx \otimes B}}}{1 - \abs{\avg{\sx \otimes B}}}}
  \varepsilon^{2} \,.
\end{equation}
The approximate device-independent bound can be derived by substituting
$\abs{\avg{\sx \otimes B}} \geq \sqrt{S\du{\alpha}{2}/4 - \alpha^{2}}$ and,
for $\abs{\alpha}<1$, replacing part of the result with its tangent as we did
for the general entropy bound in section~\ref{sec:main-result}.

To derive a generally useful approximation for the conditional Shannon
entropy we consider a joint probability distribution $p_{ab}$ of the form
\begin{equation}
  p_{ab} = \frac{p_{b} + \varepsilon \Delta_{ab}}{n_{\rA}}
\end{equation}
with $\sum_{a} \Delta_{ab} = 0$, i.e., such that $\sum_{a} p_{ab} =
p_{b}$. The joint entropy of this distribution is
\begin{IEEEeqnarray}{rCl}
  H(AB) &=& - \sum_{ab} p_{ab} \log_{2}(p_{ab}) \nonumber \\
  &=& - \sum_{ab} \frac{p_{b} + \varepsilon \Delta_{ab}}{n_{\rA}}
  \log_{2} \Bigro{\frac{p_{b} + \varepsilon \Delta_{ab}}{n_{\rA}}} \nonumber \\
  &=& - \sum_{ab} \frac{p_{b} + \varepsilon \Delta_{ab}}{n_{\rA}}
  \log_{2} \Bigsq{\frac{p_{b}}{n_{\rA}}
    \Bigro{1 + \varepsilon \frac{\Delta_{ab}}{p_{b}}}} \nonumber \\
  &=& - \sum_{ab} \frac{p_{b}}{n_{\rA}}
  \bigro{\log_{2}(p_{b}) - \log_{2}(n_{\rA})} \nonumber \\
  &&-\> \sum_{ab} \frac{p_{b} + \varepsilon \Delta_{ab}}{n_{\rA}}
  \log_{2} \Bigro{1 + \varepsilon \frac{\Delta_{ab}}{p_{b}}} \nonumber \\
  &=& H(B) + \log_{2}(n_{\rA}) \nonumber \\
  &&-\> \frac{1}{n_{\rA}}
  \sum_{ab} (p_{b} + \varepsilon \Delta_{ab})
  \log_{2} \Bigro{1 + \varepsilon \frac{\Delta_{ab}}{p_{b}}} \nonumber \\
  &\approx& H(B) + \log_{2}(n_{\rA}) \nonumber \\
  &&-\> \frac{1}{n_{\rA} \log(2)}
  \sum_{ab} \begin{IEEEeqnarraybox}[][t]{l}
    (p_{b} + \varepsilon \Delta_{ab}) \\
    \times \> \biggro{\varepsilon \frac{\Delta_{ab}}{p_{b}}
      - \varepsilon^{2} \frac{\Delta\du{ab}{2}}{2 p\du{b}{2}}}
  \end{IEEEeqnarraybox}
  \nonumber \\
  &\approx& H(B) + \log_{2}(n_{\rA}) \nonumber \\
  &&-\> \frac{1}{2 n_{\rA} \log(2)}
  \sum_{ab} \frac{\Delta\du{ab}{2}}{p_{b}} \, \varepsilon^{2} \,.
\end{IEEEeqnarray}
Rearranging this gives
\begin{equation}
  H(A|B) \approx \log_{2}(n_{\rA})
  - \frac{1}{2 n_{\rA} \log(2)} \sum_{ab} \frac{\Delta\du{ab}{2}}{p_{b}} \,
  \varepsilon^{2}
\end{equation}
for the conditional Shannon entropy.

In the applications we considered in this paper, Alice always has two
outcomes. In this case the distribution for $\varepsilon = 1$ is
\begin{equation}
  p_{ab} = \tfrac{1}{2} (p_{b} \pm \Delta_{b})
\end{equation}
with $\Delta_{b} = p_{+b} - p_{-b}$, and the approximation becomes
\begin{equation}
  \label{eq:HAB-approx}
  H(A|B) \approx 1
  - \frac{1}{2 \log(2)}
  \sum_{b} \frac{(p_{+b} - p_{-b})^{2}}{p_{b}} \, \varepsilon^{2}
\end{equation}
in terms of the joint distribution $p_{ab}$ \emph{before} noise preprocessing
is applied.

In the special case that the probabilities $P(ab|13)$ prior to noise
preprocessing being applied are of the form
\begin{IEEEeqnarray}{rCcCl}
  P({+}{+} | 1 3) &=& P({-}{-} | 1 3) &=& (1 - \delta) / 2 \,, \\
  P({+}{-} | 1 3) &=& P({-}{+} | 1 3) &=& \delta / 2 \,,
\end{IEEEeqnarray}
the approximation \eqref{eq:HAB-approx} gives
\begin{equation}
  H(A_{1} | B_{3})
  \approx 1 - \frac{(1 - 2 \delta)^{2}}{2 \log_{2}(2)} \varepsilon^{2} \,.
\end{equation}
This combined with the approximation for $H(\sz | \rE)$ above recovers
Eq.~(10) in \cite{ref:w2014}.

Losses turn an initial probability distribution $p(ab)$ to
\begin{equation}
  (p'_{ab}) = \begin{bmatrix}
    \eta^{2} \, p(++) & \eta^{2} \, p(+-) & \eta \bar{\eta} \, p_{\rA}(+) \\
    \eta^{2} \, p(-+) & \eta^{2} \, p(--) & \eta \bar{\eta} \, p_{\rA}(-) \\
    \eta \bar{\eta} \, p_{\rB}(+) &
    \eta \bar{\eta} \, p_{\rB}(-) & \bar{\eta}^{2}
  \end{bmatrix}
\end{equation}
where $\bar{\eta} = 1 - \eta$. Labelling the nondetection outcome `$\bot$',
the sum in \eqref{eq:HAB-approx} after binning nondetections on Alice's side
with `$+$' evaluates to
\begin{IEEEeqnarray}{rCl}
  \IEEEeqnarraymulticol{3}{l}{
    \sum_{b} \frac{(p'_{+b} + p'_{\bot b} - p'_{-b})^{2}}{p_{b}}} \nonumber \\
  \quad &=& \bar{\eta}^{2}
  + \eta \bar{\eta} \avg{A} \bigro{2 + \eta \avg{A}}
  \nonumber \\
  &&+\> \frac{\eta^{3}}{1 - \avg{B}^{2}}
  \Bigro{\avg{A}^{2} + \avg{AB}^{2} - 2 \avg{A} \avg{B} \avg{A B}} \,,
  \nonumber \\*
\end{IEEEeqnarray}
where
\begin{IEEEeqnarray}{rCl}
  \avg{A} &=& p_{\rA}(+) - p_{\rA}(-) \,, \\
  \avg{B} &=& p_{\rB}(+) - p_{\rB}(-) \,, \\
  \avg{A B} &=& p(++) - p(-+) - p(+-) + p(--) \,. \IEEEeqnarraynumspace
\end{IEEEeqnarray}
For $p(\pm \pm) = (1 \pm \cos(\theta))/2$ this gives
\begin{IEEEeqnarray}{rCl}
  \IEEEeqnarraymulticol{3}{l}{H(A_{1} | B_{3})} \nonumber \\
  \quad &\approx& 1 - \frac{1}{2 \log_{2}(2)} \Bigsq{
    \bigro{\bar{\eta} + \eta \cos(\theta)}^{2}
    + \eta^{3} \sin(\theta)^{2}} \varepsilon^{2} \,. \nonumber \\*
\end{IEEEeqnarray}

\section{Min-entropy and $I_{\alpha}^{\beta}$ Bell expression}
\label{sec:min-entropy}

The lower bound
\begin{equation}
  \label{eq:XB-Salpha-ge-1-bound}
  \abs{\avg{\sx \otimes B}} \geq \sqrt{S\du{\alpha}{2}/4 - \alpha^{2}}
\end{equation}
we derived for $\abs{\alpha} \geq 1$ and
\begin{equation}
  \label{eq:XB-Salpha-le-1-bound}
  \abs{\avg{\sx \otimes B}} \geq E_{\alpha}(S_{\alpha})
\end{equation}
for $\abs{\alpha} < 1$ in section~\ref{sec:derivation} can be used to derive
the tight bound for the min-entropy in terms of $S_{\alpha}$ as well as the
conditional von Neumann entropy. The min-entropy is defined as
\begin{equation}
  H_{\rmin}(A_{1} | \rE) = -\log_{2} \bigro{P_{\guess}(A_{1} | \rE)} \,,
\end{equation}
where the guessing probability $P_{\guess}(A_{1} | \rE)$ is defined as the
highest probability with which an eavesdropper can correctly guess the
outcome when Alice measures $A_{1}$. This is given by
\begin{equation}
  \label{eq:pguess-def}
  P_{\guess}(A_{1} | \rE) = P(A_{1} = E)
  = \frac{1}{2} + \frac{1}{2} \avg{A_{1} \otimes E}
\end{equation}
for whichever $\pm 1$-valued observable $E$ on Eve's system maximises the
right-hand side of \eqref{eq:pguess-def}.

Recalling that we identify $A_{1}$ with $\sz$, the correlation term
$\avg{A_{1} \otimes E}$ is bounded by
\begin{equation}
  \label{eq:AE-XB-bound}
  \avg{A_{1} \otimes E}^{2} + \avg{\sx \otimes B}^{2} \leq 1 \,.
\end{equation}
This is implied, for instance, by the family
\begin{equation}
  \bigro{\id - \cos(\theta) \, \sz \otimes \id_{\rB} \otimes E
    - \sin(\theta) \, \sx \otimes B \otimes \id_{\rE}}^{2} \geq 0
\end{equation}
of sum-of-squares decompositions. The inequalities \eqref{eq:AE-XB-bound} and
\eqref{eq:XB-Salpha-ge-1-bound} recover the tight upper bound
\begin{equation}
  P_{\guess}(A_{1} | \rE)
  \leq \frac{1}{2} + \frac{1}{2} \sqrt{1 + \alpha^{2} - S\du{\alpha}{2}/4}
\end{equation}
for the guessing probability derived for $\abs{\alpha} \geq 1$ in
\cite{ref:amp2012}. Taking the tangents of this upper bound and applying them
to the special case $E = \id$, together with the trivial bound
$\abs{A_{1}} \leq 1$, likewise recovers the Tsirelson bound
\begin{equation}
  \beta \avg{A_{1}} + S_{\alpha}
  \leq \left\{
    \begin{IEEEeqnarraybox}[\IEEEeqnarraystrutmode
        \IEEEeqnarraystrutsizeadd{3pt}{2pt}][c]{ll}
      2 \sqrt{(1 + \alpha^{2}) (1 + \beta^{2}/4)}
      &\text{ if } \abs{\beta} \leq 2/\abs{\alpha} \\
      \abs{\beta} + 2 \abs{\alpha}
      &\text{ if } \abs{\beta} \geq 2/\abs{\alpha}
    \end{IEEEeqnarraybox} \right.
\end{equation}
derived for the family $I_{\alpha}^{\beta} = \beta \avg{A_{1}} + S_{\alpha}$
of Bell expressions for $\abs{\alpha} \geq 1$ in \cite{ref:amp2012}.

For $\abs{\alpha} < 1$ the qubit bound on the guessing probability implied by
\eqref{eq:XB-Salpha-le-1-bound} needs to be partly replaced with one of its
tangents, as we needed to do for the conditional von Neumann entropy. The
result of doing this is
\begin{equation}
  \label{eq:pguess-a-le-1}
  P_{\guess}(A_{1} | \rE) \leq \left\{
    \begin{IEEEeqnarraybox}[\IEEEeqnarraystrutmode
        \IEEEeqnarraystrutsizeadd{8pt}{8pt}][c]{ll}
      \frac{1}{2} + \frac{1}{2} \sqrt{1 + \alpha^{2} - S\du{\alpha}{2}/4}
      &\text{ if } \abs{S_{\alpha}} \geq S_{*} \\
      1 - \frac{1}{\beta_{*}} (\abs{S_{\alpha}}/2 - 1)
      &\text{ if } \abs{S_{\alpha}} \leq S_{*}
    \end{IEEEeqnarraybox} \right.
\end{equation}
where
\begin{equation}
  S_{*} = 1 + \alpha^{2} + \sqrt{1 - \alpha^{4}}
\end{equation}
and
\begin{equation}
  \beta_{*} = \frac{2}{\alpha^{2}} \Bigro{1 - \sqrt{1 - \alpha^{4}}} \,.
\end{equation}
Taking the tangents again with $E = \id$ this time gives
\begin{equation}
  \beta \avg{A_{1}} + S_{\alpha} \leq \left\{
    \begin{IEEEeqnarraybox}[\IEEEeqnarraystrutmode
        \IEEEeqnarraystrutsizeadd{3pt}{2pt}][c]{ll}
      2 \sqrt{(1 + \alpha^{2})(1 + \beta^{2}/4)}
      &\text{ if } \abs{\beta} \leq \beta_{*} \\
      \abs{\beta} + 2
      &\text{ if } \abs{\beta} \geq \beta_{*}
    \end{IEEEeqnarraybox} \right. \,.
\end{equation}
This confirms that the quantum bound
\begin{equation}
  \label{eq:Iab-tsirelson-bound}
  I_{\alpha}^{\beta} \leq 2 \sqrt{(1 + \alpha^{2})(1 + \beta^{2}/4)}
\end{equation}
originally derived for $\abs{\alpha} \geq 1$ in \cite{ref:amp2012} also holds
for $\abs{\alpha} < 1$ as long as $\abs{\beta}$ is not too high.

\end{document}